\documentclass[%
reprint,
superscriptaddress,
amsmath,amssymb,
aps,
]{revtex4-1}

\usepackage{graphicx}
\usepackage{bm}
\usepackage[dvipsnames]{xcolor}
\usepackage{tabularx}
\newcolumntype{Y}{>{\centering\arraybackslash}X}

\newcommand{\efield}{{\mathcal{E}}}

\newlength{\figwidth}
\begin{document}

\title{Resonances in non-universal dipolar collisions}
\author{Tijs Karman}
\email{t.karman@science.ru.nl}
\affiliation{Radboud University, Institute for Molecules and Materials, Heijendaalseweg 135, 6525 AJ Nijmegen, the Netherlands}
\date{\today}
\begin{abstract}
Scattering resonances due to the dipole-dipole interaction between ultracold molecules, induced by static or microwave fields, are studied theoretically.
We develop a method for coupled-channel calculations that can efficiently impose many short-range boundary conditions,
defined by a short-range phase shift and loss probability as in quantum-defect theory.
We study how resonances appear as the short-range loss probability is lowered below the universal unit probability.
This may become realizable for nonreactive ultracold molecules in blue-detuned box potentials.
\end{abstract}

\maketitle

\begin{figure}
\begin{center}
\includegraphics[width=\figwidth]{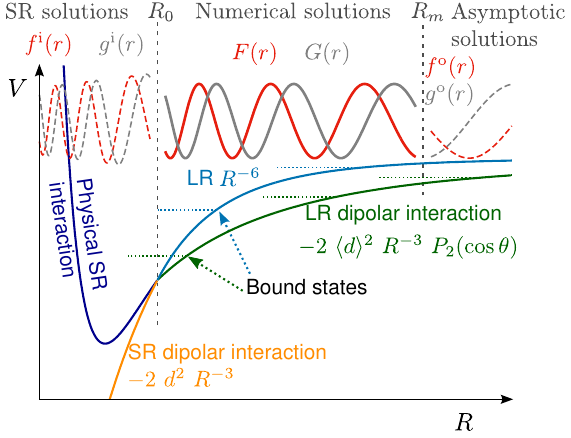}
\caption{ \label{fig:cartoon}
Sketch of the methodology and the physics explored in this work.
Scattering wavefunctions are propagated numerically between $R_0$ and $R_m$.
At asymptotically large distances, we match to the usual $S$-matrix scattering boundary conditions.
At short range, interactions between molecules dominate and the potentials become independent of the applied field.
We develop analytically the short-range boundary condition parameterized by a short-range phase shift and loss parameter, similar to the parameterization in quantum defect theory.
In our numerical calculations, the interactions are limited to dipolar interactions,
but the short-range phase shift can effectively account for the physical short-range interactions.
This method requires only one-time propagation of two independent sets of real-valued solutions,
rather than repeated propagation of complex-valued solutions initialized for each short-range phase shift and loss parameter.
\newline
At intermediate distances, where we propagate numerically,
the interaction can be varied between the rotational dispersion $R^{-6}$ interaction in the absence of fields,
and tunable long-range dipolar interactions in external fields.
Inducing long-range dipolar interactions shifts bound states to lower energy and shorter $R$.
Additional bound states appear and where these cross threshold these lead to scattering resonances that we explore in this work \cite{ticknor:05b}.
}
\end{center}
\end{figure}

Ultracold molecules have promising applications in quantum simulation \cite{micheli:06, buchler:07,pupillo:08,cooper:09, krems:09,yan:13} and computing \cite{demille:02,yelin:06,park:17,ni:18,kaufman:21}, precision measurement \cite{carr:09, krems:09,andreev:18,ho:20}.
However, ultracold molecules have been plagued by rapid collisional losses \cite{ni:08,danzl:10,takekoshi:14,molony:14,park:15,guo:16,rvachov:17,seesselberg:18,yang:19}, that occur with nearly unit probability in short-range encounters of colliding molecules \cite{idziaszek:10}.
This is sometimes referred to as ``universal loss'', since in this limit the loss rate is independent of the details of the short-range interactions that might differ from molecule to molecule,
and instead the collisional loss rate depends on the type of molecule only through a ``universal'' scaling with the mass and the strength of the van der Waals interaction \cite{idziaszek:10}.
The origin of these losses for nonreactive molecules has long been the subject of debate \cite{mayle:12}.
It has been proposed \cite{christianen:19a,christianen:19b}, and subsequently confirmed by two independent experiments \cite{gregory:20,liu:20}, that these losses are due to photochemistry initiated by the trapping laser.
This suggests it may be possible to eliminate losses by trapping using blue-detuned light that realizes uniform repulsive box potentials \cite{gaunt:13,mukherjee:17}.
For other molecules, suppression of collisional loss in the dark has so far been unsuccessful \cite{bause:21,gersema:21}.
This may be explained by an additional unforeseen loss mechanism,
or by prolonged sticking~\cite{nichols:22} due nonconservation of total angular momentum or nuclear spin states \cite{man:22b}.
The hope is that these effects can be understood and controlled,
thus realizing collisionally stable molecules.

In this work, we theoretically investigate collisions of ultracold molecules in static and microwave electric fields.
These external fields induce dipole moments in the molecules,
giving rise to dipole-dipole interactions between molecules that form the basis of most of their applications.
Tuning the strength of the long-range dipole-dipole interaction,
one can shift the energies of bound states supported by this long-range interaction.
By doing so, new bound states can appear and as these cross threshold they give rise to scattering resonances \cite{ticknor:05b}, illustrated in Fig.~\ref{fig:cartoon}.
These resonances can be observed as increased cross sections and tunable scattering length.
Resonances are absent for universal unit-probability short-range loss,
since with full absorption and no reflection at short-range, stable bound states and resonance states do not exist \cite{idziaszek:10}.
Here, we investigate how resonances emerge as the short-range loss is reduced,
as may be realizable for nonreactive molecules in repulsive box potentials.
Another avenue along which tunable long-range dipolar interactions can be realized in the absence of short-range losses is shielding \cite{gonzalez:17,karman:18d,lassabliere:18,matsuda:20,anderegg:21,schindewolf:22}.

To enable the study of non-universal molecular collisions we here develop a method where we propagate two sets of linearly independent solutions and subsequently match to the boundary conditions.
At long range, we impose the standard $S$-matrix boundary condition, whereas the short-range boundary is developed here in terms of the solutions for the short-range behavior of the $R^{-n}$ interactions employed here, for which we develop a general parameterization similar to that found in quantum defect theory \cite{mies:84a,mies:84b}.
This approach, illustrated in Fig.~\ref{fig:cartoon}, has the technical advantage of requiring only the one-time propagation of real-valued wave functions,
and yielding observables \emph{as a function of} the boundary condition.
Previous approaches required propagation of complex-valued wave functions for an \emph{individual specific} boundary condition,
which would have to be repeated for different boundary conditions \cite{wang:15}.
Unlike a previous study of dipolar scattering with non-universal short-range loss \cite{idziaszek:10b},
the present approach can be applied to multichannel scattering,
and hence also describes resonant dipolar interactions as they occur in collisions of microwave-dressed ultracold molecules \cite{yan:20,karman:21a}, for example.
Finally, since in our approach the short-range interactions and boundary conditions are field independent,
the calculations can be converged with the point at which the short-range boundary conditions are imposed, $r_\mathrm{min}$,
in contrast to previous studies that used power-law interactions with hard-wall boundary conditions at finite $r_\mathrm{min}$ to model short range boundary conditions \cite{bohn:09}.

This paper is organized as follows.
Section~\ref{sec:theory} outlines the main idea of the approach developed here.
In Sec.~\ref{sec:boundary} we develop boundary conditions, similar to those in quantum defect theory, for inverse-power-law interactions,
and Sec.~\ref{sec:wkb} gives a WKB approximation which also accounts for a finite channel energy.
In Sec.~\ref{sec:scatteringlength} we derive the zero-energy scattering length for arbitrary inverse-power-law interactions and short-range boundary conditions.
Section~\ref{sec:numerical} discusses the numerical propagation of the sets of linearly independent solutions to the Schr{\"o}dinger equation.
Section~\ref{sec:potential} describes the Hamiltonian used in numerical calculations and the long-range interactions this describes in the presence or absence of external fields.
In Sec.~\ref{sec:results} we describe numerical results for NaK molecules.
In Sec.~\ref{sec:single}, we first consider simplified single-channel calculations using the lowest adiabatic potential only.
We study the emergence of a regular series of dipolar resonances for non-universal short-range loss,
which is observable at typical experimental temperatures around 1~$\mu$K even though this is much higher than the so-called dipolar energy scale.
In Sec.~\ref{sec:multi}, we then consider multichannel scattering,
which leads to a more complex series of resonances with additional narrower features,
differences between interactions induced by microwaves or static fields,
and a transition to semiclassical scattering at higher temperature or induced dipole moment.
Concluding remarks are given in Sec.~\ref{sec:conclusions}.

\section{Theory \label{sec:theory}}

Ultracold collision dynamics is dominated by threshold effects that are sensitive to long-range behavior of the interaction potential,
which typically has a characteristic inverse-power-law behavior such as $R^{-6}$ for the van der Waals interaction between atoms, $R^{-4}$ for the polarization potential for atom-ion collisions, and $R^{-3}$ for dipolar collisions \cite{gao:08}.
At short range, the interaction potential will deviate from this inverse power law, and this too will affect the collision dynamics.

In quantum defect theory one deals with this as follows \cite{mies:84a,mies:84b,gao:98,gao:01,gao:08}.
First, one solves exactly the Schr\"odinger equation for the asymptotic $R^{-n}$ form of the interaction potential and obtains two linearly independent solutions.
If one, hypothetically, were to solve the full Schr\"odinger equation for the physical interaction,
at some large distance this interaction potential approaches its asymptotic $R^{-n}$ form and the physical wavefunction could be expressed at large $R$ as a linear combination the two solutions determined for the long-range potential.
Hence, it does not matter exactly what the microscopic short-range interaction is, as long as it solution approaches the same linear combination of long-range solutions it will produce the same observables such as scattering cross sections.
This means that the short-range interaction essentially only determines a short-range boundary condition, which can generally be parameterized by two parameters; a loss probability and phase shift.
The main simplification of quantum defect theory is the observation that this short-range boundary condition can be imposed at such short distances that the interaction is far larger than the collision energy and the centrifugal potential,
such that the boundary condition is independent of the precise collision energy and the partial wave \cite{gao:08}.
Hence, the dependence of observables on collision energy and partial wave stems entirely from the long range for which the Schr\"odinger equation is solves essentially exactly.
This leads to both a conceptual and a practical computational advantage.

The goal of this paper is to give a general description of collisions between ultracold molecules.
By polarizing molecules with external microwave or static electric fields, we can control the interactions between the molecules and switch from $R^{-6}$ van der Waals to $R^{-3}$ dipole-dipole interactions.
Our strategy is to solve the coupled-channels equations for these controllable long range interactions numerically,
which yields two independent solutions $F(R)$ and $G(R)$.
Next, we wish to impose a short-range boundary condition inspired by quantum defect theory (QDT) \cite{gao:08}.
Finally, we match the resulting wavefunction to the usual scattering boundary conditions at long range, which yields the $S$-matrix and from this all observables such as collision rates can be extracted.

In a sense, the proposed approach has already been used for example in Refs.~\cite{janssen:13,karman:18d,yan:20,karman:21a}, but only in the special case that we match to an absorbing boundary condition at short range that corresponds to complete short-range loss.
Matching to this boundary condition is somewhat simpler since it requires matching to a purely incoming wave,
which can be approximated around the matching point as $\exp(-i k R)$ where $k$ is the \emph{local} wavenumber at the matching point, which is assumed to be constant close to the matching point.
In the non-universal case, one might imagine matching to a linear combination of an incoming and reflected wave that are both defined by their local wavenumber.
However, the resulting boundary condition is then dependent on the choice of matching point, and it becomes difficult to confirm whether the numerical results are actually converged with the radial grid used in the numerical calculations.
Instead, in the spirit of quantum defect theory, we would like to define the ``short-range phase'' as $R\rightarrow 0$,
which requires knowledge of the reference solutions used for matching between $R=0$ and the matching point at some finite $R$.

The main idea of the approach for dealing with non-universal short-range boundary conditions that we develop here is that, as was done in Refs.~\cite{karman:18d,yan:20,karman:21a}, the long-range interactions between molecules can all be described microscopically by dipole-dipole interactions.
For example, even if no external fields are applied the molecules are not polarized and experience rotational van der Waals interactions is determined by the dipole-dipole interaction in second order.
Different long-range interactions induced by the presence or absence of external fields are discussed in detail in Section~\ref{sec:potential}.
Thus, if we include only pure dipole-dipole interactions, we correctly describe the long-range interactions between molecules in the presence or absence of external fields,
while the interactions deviate from the physical ones at short range.
At short range, the dipole-dipole interaction that we do account dominates over the interaction with any applied field,
making the short-range interaction effectively field independent.
This interaction then approaches $C_3 R^{-3}$ for every channel, with the coefficients determined by numerically diagonalizing the interaction matrix.
For each channel, we thus obtain a simple reference problem at short range with a power-law reference potential.
If the solutions to this problem are known to reasonable approximation, we can match to these solutions at finite $R$ while defining the boundary condition at $R=0$.
Therefore, we first study the solutions for the reference inverse-power-law potentials to which we match at short range in Sec.~\ref{sec:boundary},
and in Sec.~\ref{sec:wkb} give a WKB approximation which also accounts for a finite channel energy.
In Sec.~\ref{sec:scatteringlength} we derive the zero-energy scattering length for arbitrary inverse-power-law interactions and short-range boundary conditions, to which we will later compare numerical results.
Section~\ref{sec:numerical} discusses the numerical propagation of the sets of linearly independent solutions to the Schr{\"o}dinger equation.
Section~\ref{sec:potential} describes the Hamiltonian used in numerical calculations and the long-range interactions this describes in the presence or absence of external fields.
See Fig.~\ref{fig:cartoon} for a schematic depiction of the calculations.

We note that the boundary conditions used here are inspired by quantum defect theory developed by Gao~\cite{gao:08}.
Below, we will use a notation that is close to that of Ref.~\cite{gao:08}:
The real valued reference solutions are denoted $f_c$ and $g_c$ (instead of $f^c$ and $g^c$ in Ref~\cite{gao:08}),
whereas superscripts here will denote the approximation in which these functions are evaluated.
Incoming and outgoing waves at short range are denoted $f_i$ and $g_i$ (instead of $f^{i+}$ and $f^{i-}$ in Ref~\cite{gao:08}),
and incoming and outgoing waves at long range are denoted $f_o$ and $g_o$ (instead of $f^{o-}$ and $f^{o+}$ in Ref~\cite{gao:08}).
While these functions have the same interpretation as their counterparts in Ref~\cite{gao:08},
their definition is not exactly identical as their usage here requires flux normalization, see Sec.~\ref{sec:numerical}.
The reference solutions determined numerically by propagating the solutions to the coupled-channels equations are denoted $F$ and $G$.

\subsection{Reference solutions \label{sec:boundary}}

Consider the one-dimensional Schr\"odinger equation
\begin{align}
\left[ -\frac{\hbar^2}{2\mu} \frac{\mathrm{d}^2}{\mathrm{d}R^2} + \frac{\hbar^2 \ell(\ell+1)}{2\mu R^2} - \frac{C_n}{R^n} - E \right] \psi(R) = 0.
\end{align}
This can be cast in a dimensionless form \cite{gao:08}, introducing $r=R/\beta_n$ and $\epsilon=E/E_n$ where
the natural length and energy scales are
\begin{align}
\beta_n &= \left(2\mu C_n / \hbar^2\right)^{1/(n-2)}, \nonumber \\
E_n &= \frac{\hbar^2}{2\mu \beta_n^2}.
\end{align}
This leads to
\begin{align}
\left[ \frac{\mathrm{d}^2}{\mathrm{d}r^2} - \frac{\ell(\ell+1)}{r^2} + \frac{1}{r^n} + \epsilon \right] \psi(R) = 0.
\label{eq:seuniv}
\end{align}
For the case of van der Waals interactions, $n=6$, the solutions to this problem are known analytically \cite{gao:98}.

We denote two linearly independent solutions to Eq.~\eqref{eq:seuniv} by $f(r)$ and $g(r)$.
Arbitrary linear combinations of these solutions also satisfy Eq.~\eqref{eq:seuniv}.
Some particular choices are defined by their short-range or long-range asymptotic behavior.
In particular, the functions
\begin{align}
f_c(r) &\stackrel{r\rightarrow 0}{\simeq} \sqrt{\frac{n-2}{\pi}} r^{n/4} \nonumber \\
\times& \cos\left( \frac{2}{n-2} r^{-(n-2)/2} -\frac{2\ell+1}{n-2}\frac{\pi}{2} -\pi/4 \right), \nonumber \\ 
g_c(r) &\stackrel{r\rightarrow 0}{\simeq} \sqrt{\frac{n-2}{\pi}} r^{n/4} \nonumber \\
\times& -\sin\left( \frac{2}{n-2} r^{-(n-2)/2} -\frac{2\ell+1}{n-2}\frac{\pi}{2} -\pi/4 \right), 
\label{eq:psisr}
\end{align}
are a set of real-valued solutions with energy-independent normalization at short range.
We further define a linear combination of these solutions as
\begin{align}
f_i(r) &= \sqrt{\frac{\mu \beta\pi}{n-2}} \exp\left[-i\pi\frac{n+4\ell}{4(n-2)}\right] \times \left[ f_c(r) + i g_c(r) \right] \nonumber \\ 
&\stackrel{r\rightarrow 0}{\simeq}  \sqrt{{\mu\beta}} \, r^{n/4} \, \exp\left[-i\left(\frac{2}{n-2} r^{-(n-2)/2} \right)\right] , \nonumber \\ 
g_i(r) &= \sqrt{\frac{\mu \beta\pi}{n-2}} \exp\left[i\pi\frac{n+4\ell}{4(n-2)}\right] \times \left[ f_c(r) - i g_c(r) \right] \nonumber \\
&\stackrel{r\rightarrow 0}{\simeq}  \sqrt{{\mu\beta}} \, r^{n/4} \, \exp\left[+i\left(\frac{2}{n-2} r^{-(n-2)/2} \right)\right] ,
\label{eq:figi}
\end{align}
which have an energy and $\ell$-independent short-range normalization, and correspond to unit flux incoming and outgoing \emph{from} the origin, respectively.

\begin{figure}
\begin{center}
\includegraphics[width=\figwidth]{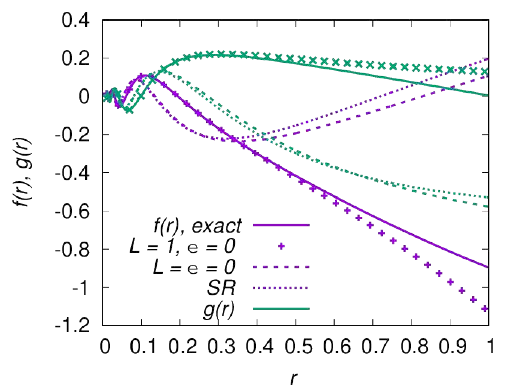}
\caption{ \label{fig:reffun}
Linearly independent solutions, $f_c(r)$ and $g_c(r)$ in purple and green, respectively, for an $r^{-3}$ potential in the energy and $\ell$-independent short-range normalization of Eq.~\eqref{eq:psisr}.
Solid lines show the exact result for $\epsilon=2$ and $\ell=1$.
Remaining lines show approximate solutions;
markers neglect the collision energy, $\epsilon=0$, Eq.~\eqref{eq:psie0},
dashed lines neglect also the centrifugal barrier, $\epsilon=\ell=0$, Eq.~\eqref{eq:psiel0},
and the dotted line shows the short range form, Eq.~\eqref{eq:psisr}.
}
\end{center}
\end{figure}

The right-hand side of Eq.~\eqref{eq:psisr} represents a short-range approximation to the reference solutions in the short-range normalization, $f_c$ and $g_c$.
For the practical application of matching numerical coupled-channels calculations, this approximation may not be as accurate as desired.
If the solutions are evaluated in an approximation that is accurate at larger $r$, this reduces the radial range over which the solutions need to be determined numerically.
The exact solutions for the $-r^{-n}$ potential, neglecting the collision energy and centrifugal kinetic energy, in the same short-range normalization are
\begin{align}
f_c^{(\ell=\epsilon=0)}(r) &=  \sqrt{r} J_{1/(n-2)}\left(\frac{2}{n-2} r^{(2-n)/2} \right), \nonumber \\ 
g_c^{(\ell=\epsilon=0)}(r) &=  -\sqrt{r} Y_{1/(n-2)}\left(\frac{2}{n-2} r^{(2-n)/2} \right),
\label{eq:psiel0}
\end{align}
where $J$ and $Y$ are the Bessel functions of the first and second kind \cite{abramowitz:64}, respectively.
Including a centrifugal barrier $\ell(\ell+1)/2r^2$ the solutions are
\begin{align}
f_c^{(\epsilon=0)}(r) &=  \sqrt{r}  J_{\nu}\left(\frac{2}{n-2} r^{(2-n)/2}\right), \nonumber \\ 
g_c^{(\epsilon=0)}(r) &=  -\sqrt{r}  Y_{\nu}\left(\frac{2}{n-2} r^{(2-n)/2}\right),
\label{eq:psie0}
\end{align}
where $\nu = (2\ell+1)/(n-2)$.
These three sets of approximations to the solutions are plotted in Fig.~\ref{fig:reffun} for $n=3$ and $\ell=1$.
The functions have the same short-range behavior by definition, but for $r>0.1$ the differences are significant.
For $\epsilon=2$, the difference between Eq.~\eqref{eq:psie0} and the exact solutions is visible only for $r>0.5$.

At long range, we define further linear combinations of $f(r)$ and $g(r)$
\begin{align}
f_{o}(r) &\stackrel{r\rightarrow \infty}{\simeq}  r \sqrt{{\mu\beta \sqrt{\epsilon}}{}} \left[j_\ell\left(\sqrt{\epsilon}r\right) - i y_\ell\left(\sqrt{\epsilon}r\right)\right] , \nonumber \\
g_{o}(r) &\stackrel{r\rightarrow \infty}{\simeq}  r \sqrt{{\mu\beta \sqrt{\epsilon}}{}} \left[j_\ell\left(\sqrt{\epsilon}r\right) + i y_\ell\left(\sqrt{\epsilon}r\right)\right] ,
\label{eq:fogo}
\end{align}
where $j$ and $y$ are the spherical Bessel functions of the first and second kind \cite{abramowitz:64}, respectively.
such that these asymptotically possess unit incoming and outgoing radial flux, respectively.

Physical potentials, $V(R)$, are not given purely by $-C_n R^{-n}$, but often approach this form asymptotically.
We denote by $r_0$ the largest distance at which the potential begins to deviate from $-C_n R^{-n}$.
For $r>r_0$, the solutions for the physical potential, $f_\mathrm{ph}$ and $g_\mathrm{ph}$, can be written as linear combinations of the independent solutions for pure $r^{-n}$ potentials discussed above.
In particular, $\psi_\mathrm{ph}(r) = f_c(r) - g_c(r) K^\mathrm{c}$, defines a short-range reactance matrix, $K^\mathrm{c}$.
Hence, the effect of an arbitrary short-range interaction potential is then completely parameterized by a short-range boundary condition.
If $r_0$ is small such that interactions at this point are strong compared to the collision energy and centrifugal kinetic energy,
the short-range boundary condition are energy and angular-momentum independent \cite{gao:01}.
Hence, the angular-momentum and energy dependence of the physical $S$-matrix arise completely due to the long-range interaction and are described by the transformation between the solutions $\{f_c(r),g_c(r)\}$ and $\{f_o(r),g_o(r)\}$.
For sufficiently simple potentials these solutions and the transformation between them are known,
leading to analytic expressions for scattering cross sections and rates as a function of the parameterized short-range interaction.
This is known as quantum defect theory \cite{mies:84a,mies:84b}.

The general short-range boundary condition can also be written as
\begin{align}
\psi \simeq \frac{(1-y)}{2\sqrt{y}} f_i(r)\exp\left(i 2\delta^s\right) + \frac{(1+y)}{2\sqrt{y}} g_i(r).
\label{eq:srbc}
\end{align}
For $0<y<1$, the boundary condition describes both an absorbed wave, $g_i(r)$, with flux towards the origin and a reflected wave, $f_i(r)$, with flux returning towards larger $r$.
The relative amplitude between the reflected and absorbed wave is given by $(1-y)/(1+y)$.
At $y=0$, the amplitudes are equal such that all flux that reaches the origin returns,
whereas at $y=1$, the amplitude of the reflected wave vanishes and all flux that reaches the origin is lost.
The denominator results from normalizing the total outgoing flux of Eq.~\eqref{eq:srbc}.
At $y=0$ the outgoing flux vanishes and normalization is not possible,
leading to the singularity in the definition.
The short-range phase shift, $\delta^s$, controls the relative phase between the absorbed and reflected waves.
The parameter $y$ determines the energy and $\ell$-insensitive probability of loss during a short-range encounter \cite{idziaszek:10}.

\subsection{WKB reference solutions including channel energy \label{sec:wkb}}

The short-range reference solutions, Eq.~\eqref{eq:psie0}, solve the short-range Schr\"odinger equation at short range for zero energy,
however, it may happen that the channel energy is not negligible at the short-range matching point.
To account for the channel energy we here consider the WKB-like solutions
\begin{align}
f_c^\mathrm{(WKB)}(r) &= \sqrt{\frac{n-2}{\pi}} q(r)^{-1/4} \nonumber \\
\times& \cos\left(p(r)  -\frac{2\ell+1}{n-2}\frac{\pi}{2} -\pi/4 \right), \nonumber \\ 
g_c^\mathrm{(WKB)}(r) &= \sqrt{\frac{n-2}{\pi}} q(r)^{-1/4} \nonumber \\
\times& -\sin\left(p(r)  -\frac{2\ell+1}{n-2}\frac{\pi}{2} -\pi/4 \right), 
\label{eq:psiwkb}
\end{align}
where $q(r)$ is the local wave number and $p(r) = \int^r q(r')\ dr'$.
In the case $\epsilon>0$ and attractive interactions
\begin{align}
q(r) &= \kappa^2 + r^{-n}, \nonumber \\
p(r) &= -r \kappa\ {}_2F_1\left(-\frac{1}{2}, -\frac{1}{n}; \frac{n-1}{n}; -\frac{r^{-n}}{\kappa^2} \right),
\label{eq:wkb1}
\end{align}
where ${}_2F_1$ denotes a hypergeometric function \cite{abramowitz:64}.
For the case $\epsilon<0$ and attractive interactions
\begin{align}
q(r) &= -\kappa^2 + r^{-n}, \nonumber \\
p(r) &= i r \kappa\ {}_2F_1\left(-\frac{1}{2}, -\frac{1}{n}; \frac{n-1}{n}; \frac{r^{-n}}{\kappa^2} \right),
\label{eq:wkb2}
\end{align}
and $\kappa=\sqrt{|\epsilon|}$.
Finally, there is the case of $\epsilon>0$ but repulsive interactions, where the solution is given by
\begin{align}
q(r) &= \kappa^2 - r^{-n}, \nonumber \\
p(r) &= -r \kappa\ {}_2F_1\left(-\frac{1}{2}, -\frac{1}{n}; \frac{n-1}{n}; \frac{r^{-n}}{\kappa^2} \right).
\label{eq:wkb3}
\end{align}
The case $\epsilon<0$ and repulsive interactions is never classically accessible and not explicitly considered here.
We note that for $\epsilon=0$, these solutions reduce to the short-range solutions discussed above.

The reference functions that account for the channel wavenumber are most relevant for calculations on scattering from excited initial states,
where lower-lying channels exist that are open at the short-range matching point and have channel energies that are significant compared to the interactions.
In this work, this is most relevant for calculations involving blue-detuned microwave dressing,
where lower-lying field dressed states occur.
In principle, these are all numerical issues;
At very short distances, the interaction will dominate every other term in the Hamiltonian, and in this case the reference solutions are known accurately.
However, this requires propagating to shorter distances, which is numerically demanding as the local wavenumber becomes high and the required step size small.
If the exact solutions for $r^{-n}$ were known, as they are for $n=6$ \cite{gao:98}, this would allow matching at larger $R_0$, reducing the numerical effort.
For the calculations reported here for static fields and red-detuned microwave dressing, where the initial channel is the lowest channel,
the attractive interaction necessarily dominates in the channels that are locally open at $R_0$,
and we numerically confirm identical results to those obtained by matching to Eq.~\eqref{eq:psie0}.

\subsection{Scattering lengths \label{sec:scatteringlength}}

We define the scattering length $a_\ell$ as the root of the asymptotic wave function for $k\rightarrow 0$,
where $k=\hbar^{-1} \sqrt{2\mu E}$ is the wavenumber.
For $n-2 \ge 2\ell+1$ this scattering length is related to the scattering phase shift as
\begin{align}
\lim_{k\rightarrow 0} -\frac{\tan\delta_\ell}{k^{2\ell+1}} = \frac{(a_\ell)^{2\ell+1}}{(2\ell+1)!!(2\ell-1)!!}.
\end{align}
Again, $a_\ell$ is the root of the asymptotic wave function $j_\ell(kR) - y_\ell(kR) \tan\delta_l$ for $k\rightarrow 0$.
We determine the scattering length from the long-range form of the zero-energy wave function, Eq.~\eqref{eq:psie0},
\begin{align}
f_c(r) &\stackrel{r\rightarrow \infty}{\simeq} c_f r^{-\ell}, \nonumber \\
g_c(r) &\stackrel{r\rightarrow \infty}{\simeq} c_g r^{\ell+1} - c_f r^{-\ell} \cot\left(\frac{2\ell+1}{n-2} \pi\right),
\end{align}
where
\begin{align}
c_f &= \left[\left(n-2\right)^{\frac{2\ell+1}{n-2}} \Gamma\left(\frac{2\ell+1}{n-2}+1\right) \right]^{-1}  \nonumber \\
c_g &= \pi^{-1}\left(n-2\right)^{\frac{2\ell+1}{n-2}} \Gamma\left(\frac{2\ell+1}{n-2}\right),
\end{align}
where $\Gamma$ is the Gamma function \cite{abramowitz:64}.
For short-range boundary conditions specified by a short-range loss probability and phase shift, $y$ and $\delta^s$, through Eq.~\eqref{eq:srbc}, we obtain
\begin{widetext}
\begin{align}
(a_\ell/\beta_n)^{2\ell+1} = i \frac{c_f}{c_g} \frac{(1-y)\left[1-i\cot\left(\frac{2\ell+1}{n-2} \pi\right) \right]\exp\left[i\left(\frac{4\ell+n}{2(n-2)}\pi +\delta^s\right)\right] + (1+y)\left[1+i\cot\left(\frac{2\ell+1}{n-2} \pi\right) \right]}{(1-y)\exp\left[i\left(\frac{4\ell+n}{2(n-2)}\pi +\delta^s \right)\right] - (1+y)}.
\label{eq:scatlen}
\end{align}
\end{widetext}
In the universal case, $y=1$, we find the scattering length is independent of $\delta^s$ and is given by $(a_\ell/\beta_n)^{2\ell+1} = \left[\cot\left(\frac{2\ell+1}{n-2} \pi\right) - i\right] c_f/c_g$.
In the special cases $n=6$ this reduces to $a_0/\beta_6 = (1-i) \pi/[2 \Gamma( {}^1/_4) \Gamma( {}^1/_4+1)]$ and $(a_1/\beta_6)^{3} =(-1-i) \pi/[2^3 \Gamma( {}^3/_4) \Gamma( {}^3/_4+1)]$.
We note this agrees with the universal scattering lengths reported by Idziaszek and Julienne \cite{idziaszek:10}.

\subsection{Numerical propagation \label{sec:numerical}}

Analytic treatments, such as that given above, are insightful and have been very successful at explaining reactive losses of ultracold molecules.
However, this approach is essentially limited to single-channel problems with potentials given by a simple analytic form, such as an inverse power law.
This includes many important cases, such as the van der Waals potential, but it cannot treat anisotropic potentials such as the dipole-dipole interaction.
Here, we first discuss the more general form of the molecule-molecule Hamiltonian we will be using,
and then discuss the numerical method used for the solution of the corresponding Schr\"odinger equation.

The Hamiltonian for the pair of colliding molecules is given by
\begin{align}
\hat{H} = -\frac{\hbar^2}{2\mu}  \frac{\mathrm{d}^2}{\mathrm{d}R^2} + \frac{\hat{L}^2}{2\mu R^2} + \hat{H}_\mathrm{mol}^{(A)} + \hat{H}_\mathrm{mol}^{(B)} + \hat{V}_\mathrm{dip-dip}(R).
\label{eq:Htot}
\end{align}
The first two terms correspond to the radial and centrifugal parts of the relative kinetic energy.
The last term represents the interaction between the two molecules,
which is here limited to the dipole-dipole interaction
\begin{align}
\hat{V}_\mathrm{dip-dip} = -\sqrt{30} R^{-3} \left[ \left[\hat{d}^{(A)} \otimes \hat{d}^{(B)}\right]^{(2)} \otimes \hat{C}^{(2)}(\hat{R}) \right]^{(0)}_{0},
\label{eq:dipdip}
\end{align}
where 
\begin{align}
\left[ \hat{A}^{(k_A)} \otimes \hat{B}^{(k_B)}\right]^{(k)}_{q} = \sum_{q_A,q_B} \langle k_A q_A k_B q_b | k q \rangle \hat{A}^{(k_A)}_{q_A} \hat{B}^{(k_B)}_{q_B}
\end{align}
is the rank-$k$ tensor product of $\hat{A}$ and $\hat{B}$,
$\hat{d}^{(X)}$ is the dipole operator for molecule $X$, see below,
the spherical components of $\hat{C}$ are Racah normalized spherical harmonics, $C_{2,q}(\hat{R})$, depending on the polar angles of the intermolecular axis,
and $\langle k_A q_A k_B q_b | k q \rangle$ is a Clebsch-Gordan coefficient.
The resulting interactions are analyzed in Sec.~\ref{sec:potential}.

The third and fourth term of Eq.~\eqref{eq:Htot} represent the monomer Hamiltonians for molecules $A$ and $B$, respectively.
The molecules are modeled as rigid rotors with a dipole moment.
The monomer Hamiltonian is given by
\begin{align}
\hat{H}_\mathrm{mol} = B_\mathrm{rot} \hat{J}^2 - \hat{d}_z \efield + \hat{H}_\mathrm{MW}.
\label{eq:monH}
\end{align}
The first term describes the rigid rotor's rotational kinetic energy, with rotational constant $B_\mathrm{rot}$.
The second term describes the Stark interaction with a static electric field along the space-fixed $z$ direction.
The third term represents the interaction with a microwave electric field
\begin{align}
\hat{H}_\mathrm{MW} = -\frac{E_\mathrm{MW}}{\sqrt{N_0}} \left( \hat{d}_\sigma \hat{a}_\sigma + \hat{d}_\sigma^{\dagger} \hat{a}_\sigma^\dagger\right) + \hbar\omega  \hat{a}_\sigma^\dagger \hat{a}_\sigma.
\end{align}
Here, $a_\sigma^\dagger$ and $a_\sigma$ are creation and annihilation operators for photons with polarization $\sigma$ and angular frequency $\omega$.
The dipole operator has spherical components $\sigma=0,\pm1$ which are related to the Cartesian components by $\hat{d}_0 = \hat{d}_z$ and $\hat{d}_{\pm 1} = \mp \left( \hat{d}_x \pm i \hat{d}_y\right)/\sqrt{2}$, corresponding to polarizations $\pi$ and $\sigma^\pm$.

In coupled-channels calculations,
one introduces a basis set for all coordinates except the radial coordinate.
Here, we use basis functions of the form
\begin{align}
|\tilde{j}_A m_{A}\rangle|\tilde{j}_B m_{B}\rangle|\ell m_\ell\rangle|N_\mathrm{MW}\rangle,
\end{align}
which describe the rotational state for both molecules,
the relative angular momentum of the colliding molecules $\ell$,
and the microwave photon number $N_\mathrm{MW}$.
The functions $|\tilde{j}_A m_{A}\rangle$ are obtained as eigenstates of the molecule in a static external field, \emph{i.e.},\ $\tilde{j}$ correlates  to the rotational angular momentum at low $\efield$.
Spherical harmonics up to $n=3$ were included in order to calculate these eigenstates.
These channel functions are adapted to permutation of identical molecules as is described in Ref.~\cite{karman:18d}.
The basis sets are truncated by including only functions with $\tilde{j}=0,1,2$ and 3, $\ell$ even or odd integers up to 30, and $N_\mathrm{MW}=N_0, N_0-1,N_0-2$.
For $\ell>6$, only $\tilde{j}=0$ and 1 had to be included.

Expanding the scattering wave function in the channel basis introduced above
\begin{align}
\Psi_j(R) = \frac{1}{R} \sum_i |\phi_i\rangle \Phi_{i,j}(R),
\end{align}
yields a set of coupled differential equations
\begin{align}
\frac{\mathrm{d}^2}{\mathrm{d}R^2} \bm{\Phi} &= \bm{W} \bm{\Phi}, \nonumber \\
W_{i,j}(R) &= 2\mu \langle \phi_i | \frac{\hbar^2 \hat{L}^2}{2\mu R^2} + \hat{H}^{(A)} + \hat{H}^{(B)} + \hat{V}(R) - E |\phi_j\rangle.
\end{align}
The coupled equations are typically solved numerically by discretizing the radial coordinate into grid points $R_0, R_1, R_2, \ldots, R_m$,
initializing the wave function using a short-range boundary condition,
and propagating the solution to large $R$.
At the last grid points, the solution is then matched to the $S$-matrix boundary condition,
which yields the $S$ matrix and thereby scattering lengths, cross sections, and rate coefficients.
For numerical stability one typically propagates a derived property that is insensitive to exponential scaling of the amplitudes of locally closed channels,
such as the log-derivative matrix $\bm{Y}_i \bm{\Phi}_i = \bm{\Phi}'_i$ or the renormalized $Q$-matrix, $\bm{Q}_i \bm{\Phi}_i = \bm{\Phi}_{i-1}$, but the principle remains the same.

\subsection{Imposing the boundary conditions \label{sec:boundaryconditions}}

In practice, the boundary condition used to initialize the wave function is often a hard wall at the first grid point, $\bm{\Phi}_0 = \bm{0}$,
which is typically chosen at such short $R$ that the potential has become highly repulsive and the wavefunction is exponentially small.
The hard-wall boundary condition could also be imposed at any desired $R$,
which does not lead to calculations converged with $R_0$, but this approach has been used to effectively explore different short-range boundary conditions in previous studies of the dipole-dipole interaction \cite{bohn:09}.
It is also straightforward to initialize the short-range wave function using the QDT boundary conditions, parameterized by $y$ and $\delta^s$, considered here using Eq.~\eqref{eq:srbc}.
This approach has been taken previously in Refs.~\cite{wang:15,croft:20}.
This approach requires propagation of a complex-valued wave function, rather than a real-valued one.
Furthermore, exploring various boundary conditions, \emph{i.e.},\ values of $y$ and $\delta^s$, then requires repeating the full calculation many times.

As an alternative, we use the renormalized Numerov algorithm of Ref.~\cite{janssen:12,janssen:13}.
This method yields two linearly independent sets of real-valued solutions,
one defined by $\bm{F}_0 = \bm{0}$ and $\bm{F}_m = \bm{1}$,
the other by $\bm{G}_0 = \bm{1}$ and $\bm{G}_m = \bm{0}$.
Subsequently, any desired boundary condition can be imposed.
The particular boundary condition chosen is that at long range there is unit incoming flux in the initial state as well as outgoing flux in the asymptotically open channels, defined by the $S$-matrix,
while at short range flux escapes into ``reactive'' channels
\begin{align}
\bm{\Phi}_m &= \bm{I}_m + \bm{O}_m \bm{S}^{(m)}, \nonumber \\
\bm{\Phi}_0 &= \bm{O}_0 \bm{S}^{(0)}.
\end{align}
The matrices $\bm{I}_m$ and $\bm{O}_m$ are diagonal and their diagonal elements contain asymptotic incoming and outgoing solutions, $f_o(r)$ and $g_o(r)$, see Eq.~\eqref{eq:fogo}.
We assume the short-range solutions uncouple in the local adiabatic basis,
\emph{i.e.},\ $\bm{U}_0^\dagger \bm{O}_0$ is diagonal, where $\bm{U}_0$ is the unitary transformation between the channel and adiabatic representation at $R_0$.
As explained in more detail in the following paragraph, the diagonal elements of $\bm{U}_0^\dagger \bm{O}_0$ are given by Eq.~\eqref{eq:srbc} repeated here for clarity
\begin{align}
\psi \simeq \frac{(1-y)}{2\sqrt{y}} f_i(r)\exp\left(i2\delta^s\right) + \frac{(1+y)}{2\sqrt{y}} g_i(r).
\label{eq:srbc2}
\end{align}
The matrix $\bm{O}_0$ itself is then obtained by transforming back to the primitive basis.
Explicitly, we obtain the inelastic and reactive blocks of the $S$-matrix as
\begin{widetext}
\begin{align}
\bm{S}^{(m)} &= -\left[ \bm{F}_{m-1} \bm{O}_m - \bm{O}_{m-1} - \bm{G}_{m-1} \bm{O}_0 \left( \bm{G}_1 \bm{O}_0 - \bm{O}_1\right)^{-1} \bm{F}_1 \bm{O}_m \right]^{-1} \left[ \bm{F}_{m-1} \bm{I}_m - \bm{I}_{m-1} - \bm{G}_{m-1} \bm{O}_0 \left( \bm{G}_1 \bm{O}_0 - \bm{O}_1\right)^{-1} \bm{F}_1 \bm{I}_m \right], \nonumber \\
\bm{S}^{(0)} &= -\left(\bm{G}_1\bm{O}_0-\bm{O}_1\right)^{-1} \bm{F}_1 \left( \bm{I}_m + \bm{O}_m \bm{S}^{(m)} \right).
\end{align}
\end{widetext}
We note that the inelastic $S$-matrix, $\bm{S}^{(m)}$ is given in the asymptotic basis that diagonalizes the asymptotic Hamiltonian, which here coincides with the primitive channel basis,
whereas the the rows of the reactive $S$-matrix, $\bm{S}^{(0)}$, correspond to the locally adiabatic channels at short range.
For reactive channels, the square matrix elements of $\bm{S}^{(0)}$ can be interpreted as the probability for capture in a particular locally adiabatic channel at short range.
The columns of the combined inelastic and reactive $S$-matrix, restricted to open asymptotic and reactive channels, are orthonormal if $\bm{I}_m$, $\bm{O}_m$, and $\bm{O}_0$ are all flux-normalized.

As noted above, the matrix of short-range solutions is diagonal in the locally adiabatic basis, and here we summarize which expressions are used for its diagonal elements.
For each adiabat, we determine the channel energy, angular momentum and interaction strength, $E$, $\ell$, and $C_n$.
These are determined by transforming the asymptotic Hamiltonian, centrifugal barrier, and dipole-dipole interaction to the locally adiabatic basis, respectively.
For adiabats that are locally closed with local wavenumber $k$, we match to
\begin{align}
o(R) = \exp(k R).
\end{align}
For locally open short-range adiabats we use Eq.~\eqref{eq:srbc2}, but we distinguish two approaches for numerically evaluating the incoming and reflected short-range waves, $f_i$ and $g_i$.
The first option is to match to neglect the channel energy, in which approximation the exact solutions are given in Eq.~\eqref{eq:psie0}.
The second option is to include the channel energy and evaluate the solutions approximately using WKB,
inserting Eq.~\eqref{eq:psiwkb} into Eqs.~\eqref{eq:figi} and \eqref{eq:srbc}.
Which expression is used for the WKB amplitude and phase --- Eqs.~\eqref{eq:wkb1}, \eqref{eq:wkb2}, or~\eqref{eq:wkb3} --- depends on the sign of the local interaction and channel energy.
We note that we consider the case of locally open channels with repulsive interactions
to constitute \emph{nonreactive} channels, and we match directly to the locally sine or cosine-like solution.
We find no dependence on this local phase.
If there would be a dependence, the correct linear combination of the two solutions could be determined from the WKB connection formulae at the inner classical turning point.
Here, we do not go into this detail.
We note this hypothetical situation cannot arise in calculations completely converged with $R_0$;
if $R_0$ is small enough that the interaction dominates each adiabat, all locally accessible adiabats correspond to attractive interactions.
Since the exact zero-energy reference functions and the WKB reference functions have the same short-range behavior, the two approaches should yield the same results when calculations are converged with $R_0$.
On the difference between the two approaches outlined here we note the WKB treatment is more appropriate for calculations involving blue-detuned microwaves,
as here the initial state is not the lowest channel such that open channels with channel energies exceeding the interaction strength can occur.
For all other calculations, we obtain excellent agreement between results using the WKB reference solutions and the exact solutions neglecting the channel energy.

The Numerov algorithm of Ref.~\cite{janssen:12,janssen:13} has previously already been applied to impose capture boundary conditions based on the local wave number in each adiabatic channel.
Matching to plane waves depending on the local channel wave number, however, one cannot define the phase at short range as this would depend on the local wavenumber at shorter $R$, which is not accounted for.
Therefore, using this method, one can only match to fully-absorbing universal capture boundary conditions ($y=1$) where the results are independent on the short-range phase.
Using the method presented here, however, we match to the analytic solutions for the short-range interaction which account for the local wavenumber at short $R$ exactly.
Hence, this method enables a consistent definition of the short-range phase,
and matching to boundary conditions for arbitrary $y$ and $\delta^s$, \emph{i.e.}, channel and energy-independent short-range parameters.

Cross sections can be computed from the matrix $\bm{T}^{(m)} = \bm{S}^{(m)} - \bm{1}$ as
\begin{align}
\sigma_{i\rightarrow f} = \frac{2\pi}{k^2} \sum_{\ell,m_\ell,\ell',m'_\ell} |T^{(m)}_{f,\ell',m'_\ell;\ i,\ell,m_\ell}|^2,
\end{align}
where $k$ is the channel wavenumber, and $i$ and $f$ are initial and final states, and the factor of two is applicable only for indistinguishable molecules in identical initial states.
Elastic cross sections refer to $f=i$,
whereas cross sections for $f\neq i$ are referred to as inelastic.
The cross section for reaching short range, or reactive loss, is given by
\begin{align}
\sigma_\mathrm{SR} = \frac{2\pi}{k^2} \sum_{r,\ell,m_\ell} |S^{(0)}_{r;\ i,\ell,m_\ell}|^2,
\end{align}
where $r$ enumerates the ``reactive'' locally adiabatic short-range channels. 
Thermal rate coefficients are calculated by averaging the velocity times the cross sections over a Maxwell-Boltzmann distribution
\begin{align}
\beta = (k_B T)^{-3/2} 2\sqrt{\frac{2}{\pi \mu}} \int_0^\infty E \sigma(E) \exp\left(-\frac{E}{k_BT}\right)\ dE,
\end{align}
where $k_B$ is the Boltzmann constant.
The thermal average, where applicable, is computed by numerical integration using a logarithmically spaced discrete grid of energies ranging from at least a factor of 10 below the stated temperature to a factor of 50 above it.

\subsection{Interaction potentials \label{sec:potential}}

In this work, the interaction between the molecules is limited to the dipole-dipole interaction, which is dominant at long range.
However, the molecules' dipole moments are attached to their bond axis.
This axis becomes aligned or oriented in space only in external fields, and so these can be used to control the intermolecular interaction.

%C6
In the absence of external fields, the ground molecular state is just the rotational ground state, $|\tilde{j}=0,m=0\rangle = |j=0,m=0\rangle$.
This eigenstate has zero dipole moment, $\langle\hat{d}\rangle = 0$, and so no first-order interaction.
However, the dipole-dipole interaction does couple to the rotationally excited state.
Treating this in second-order perturbation theory yields an isotropic van der Waals potential $V(R) = -C_6 R^{-6}$ with $C_6=d^4/6B_\mathrm{rot}$ where $B_\mathrm{rot}$ is the rotational constant.
We can define characteristic length and energy scales for this potential as
\begin{alignat}{2}
\beta_6 &= \left(2\mu C_6 / \hbar^2 \right)^{1/4} &&\approx 490~a_0, \nonumber \\
E_6 &= \left(2\mu \beta_6^2 \right)^{-1} &&\approx 11~\mu\mathrm{K},
\end{alignat}
where the numerical values are given for NaK molecules.

%C3
If an external static field is applied, the lowest molecular eigenstate $|\tilde{j}=0,m=0\rangle$ will become polarized along the field direction.
At high fields, the induced dipole moment will saturate at the magnitude of the body-fixed dipole moment.
This leads to a first-order interaction for a pair of molecules in their lowest state
\begin{align}
V(R) = -2\langle d\rangle^2/R^3 P_2(\cos\theta),
\label{eq:dipdipsimple}
\end{align}
where $\theta$ is the angle between the intermolecular axis and the electric field direction.
Because the interaction is anisotropic, it does not strictly speaking follow Eq.~\eqref{eq:seuniv},
but its multichannel equivalent can still be made universal using the characteristic length and energy scales
\begin{alignat}{2}
\beta_3 &= 2\mu d^2 &&\approx 1.3\cdot 10^5~a_0, \nonumber \\
E_3 &= \left(2\mu \beta_3^2 \right)^{-1} &&\approx 160~\mathrm{pK}.
\end{alignat}
We note that the values are given for the limiting value of the dipole moment of NaK.
For smaller static fields, a smaller fraction of the total dipole moment will be induced,
corresponding to a shorter characteristic length and a larger characteristic energy.

%microwave case
Microwave electric fields induce rapidly oscillating or rotating dipole moments in the molecules.
Time averaging over this fast rotation one obtains a first-order dipole-dipole interaction,
and we define an ``equivalent dipole moment'' by equating the first-order interaction to Eq.~\eqref{eq:dipdipsimple}.
The maximum dipole moment is $\langle d \rangle = d/\sqrt{6}$ for linear $\pi$ polarization and $\langle d\rangle = i d /2\sqrt{3}$ for circular $\sigma^\pm$ polarization.
We note the imaginary equivalent dipole moment for circular polarization reflects sign reversal of the dipole-dipole interaction.
This maximum dipole moment is induced on resonance, $\Delta=0$,
and decreases with the ratio of detuning and Rabi frequency, $\Delta/\Omega$ \cite{karman:21a}.

%C4
For bosonic molecules, the lowest adiabatic channel asymptotically corresponds to $\ell=0$.
Taking the expectation value of the anisotropic dipole-dipole interaction $\sim P_2(\cos\theta)$ leads to zero first-order interaction in the lowest adiabat.
However, the dipole-dipole interaction does couple this channel to $\ell=2$, which lies above the $\ell=0$ channel by $6\hbar^2/2\mu R^2$.
Treating this coupling in second order leads to an isotropic $C_4 R^{-4}$ potential with $C_4 = 4/15 \mu d^4$.
\begin{alignat}{2}
\beta_4 &= \left(2\mu C_4 / \hbar^2 \right)^{1/2} &&\approx 2.0\cdot 10^4~a_0, \nonumber \\
E_4 &= \left(2\mu \beta_4^2 \right)^{-1} &&\approx 1.2~\mathrm{nK}.
\end{alignat}

%very short range
At very short range, the dipole-dipole interaction will dominate over centrifugal kinetic energy and even the monomer Hamiltonian,
meaning that each adiabatic potential will behave as $C_3 R^{-3}$.
The point at which this occurs is roughly where the dipole-dipole interaction is comparable to the rotational constant
\begin{alignat}{2}
\beta_B &= \left(\frac{d^2}{4\pi\epsilon_0 B}\right)^{1/3} &&\approx 140~a_0.
\label{eq:betaB}
\end{alignat}
We use this to match the solutions to the short-range $R^{-3}$ potential at $R_0=20~a_0$ for all calculations,
regardless the behavior of the potential at long range.

%note on corrections to the potential. 
Higher multipole moments, higher-order long-range interactions, and complete modifications of the interaction at short range exist, but are not included here.
These may well affect the physical potential at the matching point, but as we will see, the dynamics is completely determined by the long-range potential.
Although excluded explicitly from the calculation,
their effects are then effectively modeled by the short-range phase shift, $\delta^s$.
Of the long-range interactions that determine the dynamics, the rotational van der Waals interaction with $\beta_6 \approx 490~a_0$ is the shortest ranged.
At these distances, the next electrostatic interaction that we excluded, the dipole-quadrupole interaction, is weaker than the dipole-dipole interaction by about two orders of magnitude.
We have also excluded the electronic van der Waals interaction, which is weaker than the rotational contribution by a factor 60.

\section{Results \label{sec:results}}
\subsection{Single Adiabat model \label{sec:single}}

\begin{figure}
\begin{center}
\includegraphics[width=\figwidth]{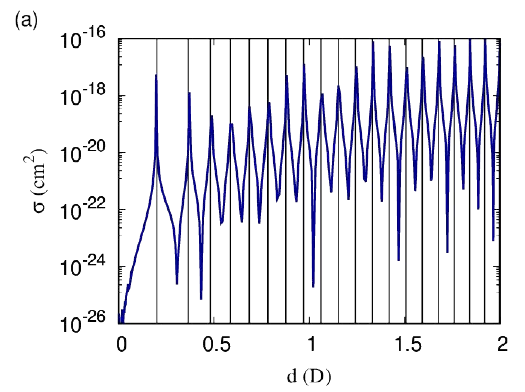}
\includegraphics[width=\figwidth]{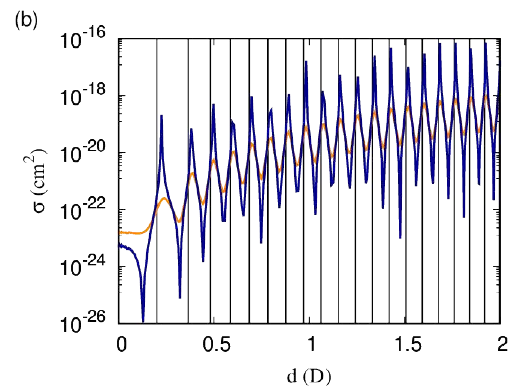}
\caption{ \label{fig:hwb}
Elastic cross section as a function of the induced dipole moment.
Parameters correspond to bosonic NaK,
the collision energy is set to $10$~pK,
and only the lowest adiabatic potential curve is used.
Vertical lines indicate the positions of resonances from a WKB estimate.
Panel (a) has been obtained with a hard-wall boundary condition imposed at $R=20~a_0$,
and panel (b) shows in blue the nonreactive QDT boundary condition with $y=0$ and $\delta^s=3\pi/4$,
and in yellow a reactive QDT boundary condition with $y=0.5$ and $\delta^s =3\pi/4$.
}
\end{center}
\end{figure}

Dipolar collisions -- due to the anisotropy of the interactions -- involve multiple partial waves,
and hence are described by multi-channel scattering.
To simplify the analysis, we first consider single-channel collisions on the lowest diabatic potential, $V_0(R)$,
which is the lowest eigenvalue of the Hamiltonian excluding radial kinetic energy as a function of the intermolecular distance, $R$.
Figure~\ref{fig:hwb} shows elastic cross sections for collisions between bosonic NaK molecules obtained using this simplified model as a function of the dipole moment induced by applying a static electric field.
Panel~\ref{fig:hwb}(a) shows results obtained with a hard-wall boundary condition imposed at $R_0=20~a_0$.
The vertical lines indicate the resonance positions, $d_m$,
estimated using the WKB approximation to the appearance of an additional bound states,
\begin{align}
\int_{R_0}^\infty \sqrt{-2\mu V_0(R,d_m) }~\mathrm{d}R = m \pi,
\label{eq:wkbint}
\end{align}
where $V_0$ is the lowest adiabatic potential at a given induced dipole moment.
Note that WKB quantization may require an additional phase shift, which is omitted since we are simply interested in estimating the number of resonances.
We note that the WKB estimate of the total number of bound states does not converge as $R_0\rightarrow 0$ for inverse-power-law potentials,
but the number of additional states supported by the external-field-induced interactions converges with $R_0\ll \beta_B$, Eq.~\eqref{eq:betaB}, where the dipole-dipole interaction between the molecules dominates over the interaction with the external field such that $V_0$ becomes independent of the applied field.
These results are similar to those reported by Bohn, Cavagnero and Ticknor \cite{bohn:09},
except that in that work the hard-wall boundary condition was used to effectively model short-range physics in a calculation that explicitly accounts only for a long-range $r^{-3}$ potential.
Hence, the number of resonances supported by the field-dependent long-range $r^{-3}$ potential, Eq.~\eqref{eq:wkbint}, is dependent on the somewhat arbitrary choice of $R_0$ which simultaneously determines the short-range phase.
In the approach taken here, the interaction naturally becomes field independent at short range where the dipole-dipole interaction dominates,
such that the position of the hard-wall boundary condition, $R_0$, determines the short-range phase and hence the position of the resonances,
but not the number of resonances induced by applying an external field.
Fig.~\ref{fig:hwb}(b) shows resonances in the elastic cross section obtained for a nonreactive QDT boundary condition, $y=0$ and $\delta^s=3\pi/4$, Eq.~\eqref{eq:srbc}.
We observe that the density of resonances again matches with the WKB estimate when both calculations are converged with $R_0$.
However, there exists a shift in position of the resonances between calculations based on the hard-wall boundary condition -- where the short-range phase is set by $R_0$ -- and QDT-like boundary condition,
where the short-range phase is set explicitly as a parameter and is independent of the matching point $R_0$.
Figure~\ref{fig:hwb}(b) also shows in yellow a broadening of the resonances caused by short-range loss $y=0.5$, whereas the resonance position controlled by $\delta^s=3\pi/4$ is unchanged.
In the absence of loss, the contrast may be determined by the grid resolution.

\begin{figure}
\begin{center}
\includegraphics[width=\figwidth]{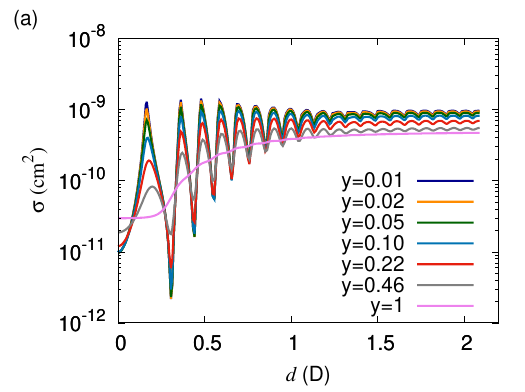}
\includegraphics[width=\figwidth]{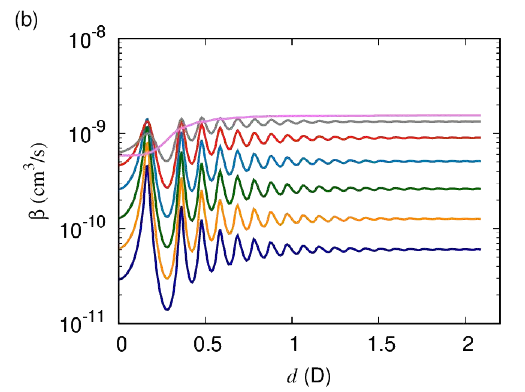}
\caption{ \label{fig:nonuniv}
Elastic cross sections (a) and short-range loss rate (b) as a function of the induced dipole moment.
Parameters correspond to bosonic NaK at a temperature of $1~\mu$K,
and only the lowest adiabatic potential curve is used.
Different curves correspond to different values of $y$ logarithmically scaled between 1 and 0.01,
and $\delta^s =3\pi/4$ throughout.
For universal loss, $y=1$, a smooth increase of both the elastic cross section and loss rate are observed with induced dipole moment,
which increases the range of the interaction.
For non-universal losses as high as $y\approx0.5$ a series of resonances in the cross sections and loss rates emerges.
For losses below about $y=0.1$, the elastic cross section converges and the loss monotonically decreases with decreasing $y$, but is otherwise independent of $y$.
}
\end{center}
\end{figure}

Next, we examine the dependence on the short-range loss parameter, $y$.
Figure~\ref{fig:nonuniv} shows elastic cross sections and short-range loss rates as a function of the induced dipole moment.
These are obtained for the simplified single-channel model that uses only the lowest adiabatic potential.
Parameters correspond to bosonic NaK at a temperature of $1~\mu$K.
Different curves correspond to different values of $y$ between 1 and 0.01,
and fixed $\delta^s=3\pi/4$ throughout.
For universal loss, $y=1$, a smooth increase of both the elastic cross section and loss rate are observed with induced dipole moment,
which increases the range of the interaction.
At large induced dipole moment this curve flattens, which is an artifact of the single-channel model, as we will see below.
For non-universal losses as high as $y\approx0.5$ a series of resonances in the cross sections and loss rates emerges.
For losses below about $y=0.1$, the elastic cross section converges and the loss monotonically decreases with decreasing $y$, but is otherwise independent of $y$.
At higher induced dipole moment these resonances become less clearly observable.

\begin{figure}
\begin{center}
\includegraphics[width=\figwidth]{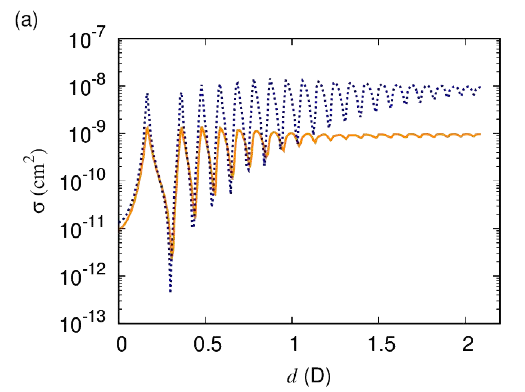}
\includegraphics[width=\figwidth]{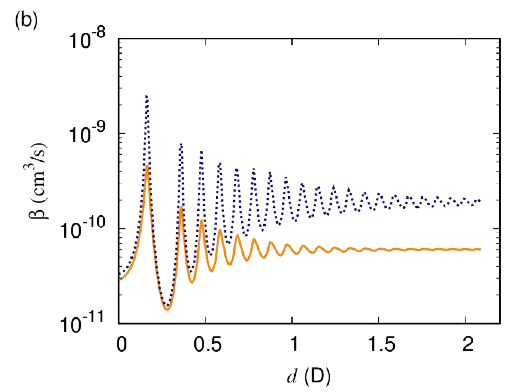}
\caption{ \label{fig:nonuniv100nK}
Elastic cross sections (a) and short-range loss rate (b) as a function of the induced dipole moment for scattering on the lowest adiabatic potential.
The solid orange and dotted blue curves correspond to temperatures $1~\mu$K and $0.1~\mu$K, respectively.
As a larger moment is induced, the characteristic energy scale of the dipole-dipole interaction decreases,
and as this energy drops below the thermal energy, the resonances become washed out.
As the temperature is lowered, the series of resonances becomes more clearly observable at higher induced dipole moment.
At experimentally realizable temperatures a significant part of the series is observable.
}
\end{center}
\end{figure}

We consider the dependence on temperature by comparing elastic cross sections and short-range loss rates at $1~\mu$K and $0.1~\mu$K.
These are compared in Fig.~\ref{fig:nonuniv100nK} for $y=0.01$ and $\delta^s =3\pi/4$,
At the lower temperature, the resonances are more clearly observable.
As a larger moment is induced, the characteristic energy scale of the dipole-dipole interaction decreases,
and as this energy drops below the thermal energy, the resonances become washed out.
As the temperature is lowered, the series of resonances becomes more clearly observable at higher induced dipole moment.
At experimentally realizable temperatures around 1~$\mu$K, a significant part of the series is observable.

\begin{figure}
\begin{center}
\includegraphics[width=\figwidth]{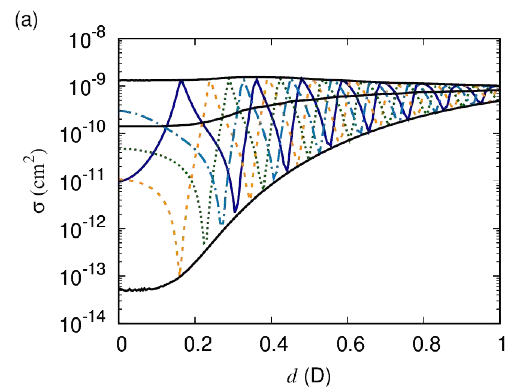}
\includegraphics[width=\figwidth]{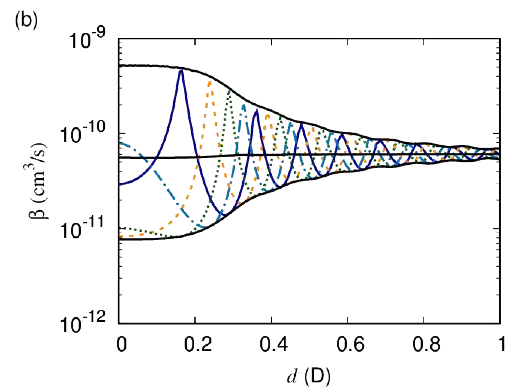}
\caption{ \label{fig:phase}
Elastic cross sections (a) and short-range loss rate (b) as a function of the induced dipole moment.
Different curves correspond to different values of $\delta^s$,
whereas $y=0.01$ throughout.
The short-range phase shift determines the positions of the resonances,
as well as the dynamics as vanishing induced dipole moment.
}
\end{center}
\end{figure}

Finally, we inspect the dependence on the short-range phase shift.
Figure~\ref{fig:phase} shows elastic cross sections and short-range loss rates for various short-range phase shifts, $\delta^s$, for fixed $y=0.01$.
The phase shift determines the position of the resonances as well as the cross sections at zero induced dipole moment.
The black lines indicate the maximum, mean, and minimum over the short-range phase, $\delta^s$, respectively.
At larger induced dipole moment the resonances become less pronounced and the dependence on the short-range phase shift decreases,
such that the envelope of possible cross sections and loss rates, for fixed $y$, becomes more restrictive.

\begin{figure}
\begin{center}
\includegraphics[width=\figwidth]{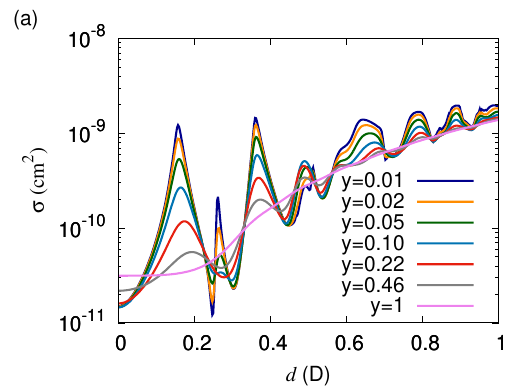}
\includegraphics[width=\figwidth]{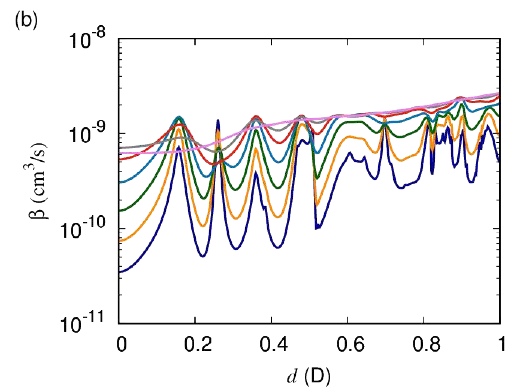}
\caption{ \label{fig:multichannel}
Elastic cross sections (a) and short-range loss rate (b) as a function of the induced dipole moment at a temperature of $1~\mu$K,
from multichannel coupled-channels calculations.
Different curves correspond to different values of $y$ logarithmically spaced between 1 and 0.01,
and $\delta^s =3\pi/4$ throughout.
}
\end{center}
\end{figure}

\subsection{Multichannel scattering \label{sec:multi}}

After examining the simplified single-adiabat model considered above,
we consider multi-channel scattering due to anisotropic dipolar interactions.
Figure~\ref{fig:multichannel} shows elastic cross sections and short-range loss rates for collisions of bosonic NaK molecules at a temperature of $1~\mu$K.
Qualitatively, these results are similar to those of the single-channel model, Fig.~\ref{fig:nonuniv}.
The cross sections increase with dipole moment, and a series of resonances emerges for non-universal loss that is clearly observable already at losses as high as $y=0.5$.
The elastic cross sections converge for $y\le 0.1$, whereas the loss rate continues to decrease monotonically with decreasing $y$.
In addition, the higher partial waves give rise to a continuing increase of the cross section with dipole moment, which increases with the length scale of the dipole-dipole interaction, 
that was absent in the single-channel model.
The higher partial waves also contribute additional narrow resonances that appear only for smaller short-range loss parameters, $y$.

\begin{figure}
\begin{center}
\includegraphics[width=\figwidth]{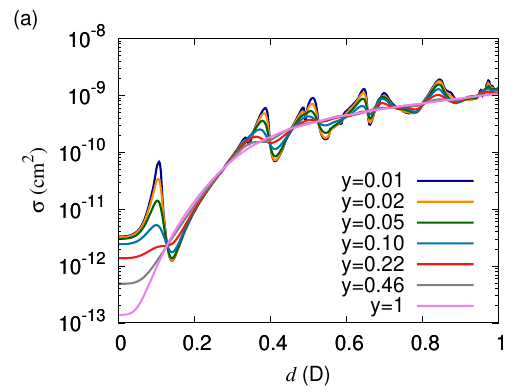}
\includegraphics[width=\figwidth]{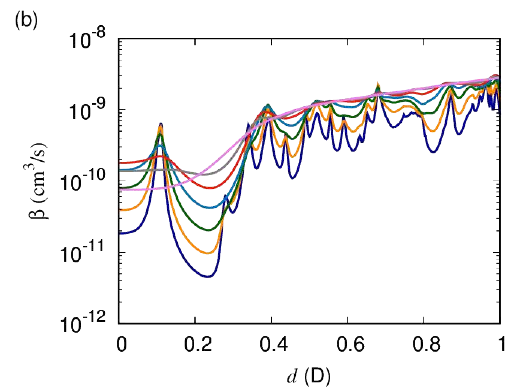}
\caption{ \label{fig:multichannel_fermion}
Elastic cross sections (a) and short-range loss rate (b) as a function of the induced dipole moment at a temperature of $1~\mu$K,
from multichannel coupled-channels calculations for Fermionic NaK.
Different curves correspond to different values of $y$ between 1 and 0.01,
and $\delta^s=3\pi/4$ throughout.
}
\end{center}
\end{figure}

Elastic cross sections and short-range loss rates for collisions of \emph{fermionic} NaK molecules are shown in Fig.~\ref{fig:multichannel_fermion}.
These were obtained from multichannel coupled-channels calculations at a temperature of $1~\mu$K.
Compared to the bosonic case, the increase of the cross section and loss rate from that at zero induced moment is much more dramatic.
At zero dipole moment, the cross sections are suppressed by the centrifugal barrier, leading to an elastic cross section that scales as $T^2$ and an inelastic rate that scales as $T$.
At ultracold temperatures, these become much smaller than the cross sections in the case of dipolar scattering.
Otherwise, the main features are similar to those observed for scattering of bosonic molecules:
we find a series of resonances emerges for non-universal loss that should be observable already at losses as large as $y=0.5$ and achievable temperatures below 1~$\mu$K.

\begin{figure}
\begin{center}
\includegraphics[width=\figwidth]{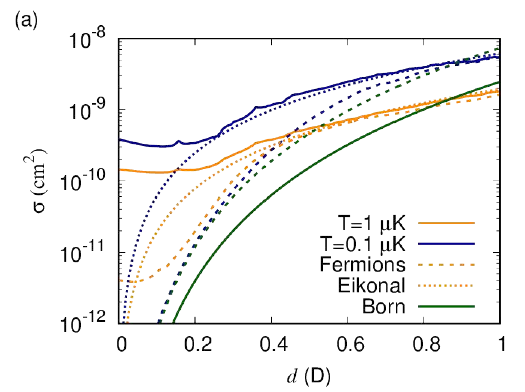}
\includegraphics[width=\figwidth]{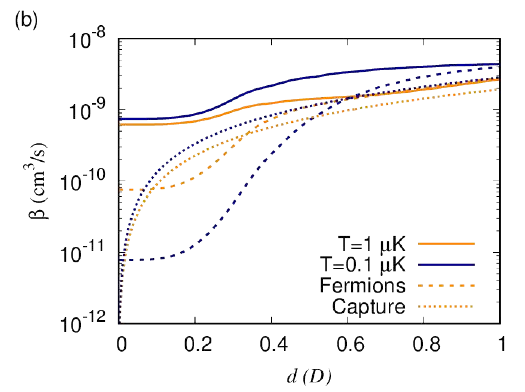}
\caption{ \label{fig:multichannel_T2} Elastic cross sections (a) and short-range loss rate (b) as a function of the induced dipole moment for $T=0.1$ and 1~$\mu$K in orange and blue, respectively.
Elastic cross sections are given for $y=0$ and averaged over $\delta^s$,
whereas short-range loss rates are given for $y=1$ and are independent of $\delta^s$.
Results for bosons and fermions are shown as solid and dashed lines, respectively.
Dotted lines show analytical results in the semiclassical Eikonal approximation for the elastic cross section,
and the Langevin rate for short-range loss.
}
\end{center}
\end{figure}

Next, we revisit the temperature dependence in the multichannel case.
Figure~\ref{fig:multichannel_T2} shows elastic cross sections and short-range loss rates for bosonic and fermionic NaK molecules as a function of the induced dipole moment for $T=0.1$~$\mu$K and 1~$\mu$K.
Here, elastic cross sections are given for $y=0$ and averaged over $\delta^s$,
whereas short-range loss rates are given for $y=1$ and are independent of $\delta^s$.
This has removed resonance structures in the cross sections that are dependent on the short-range phase,
which facilitates examination of the temperature and dipole moment dependence of the background.
For bosonic molecules, the field-free cross section and rate are due to $s$-wave collisions on the van der Waals potential,
and hence independent of temperature as long as we are away from resonance.
When averaged over the short-range phase shift, the resonant contribution leads to an increase for lower temperatures.
For fermionic molecules, the elastic cross section and short-range loss rate scale as $T^2$ and $T$, respectively, due to $p$-wave collisions.
At higher induced dipole moment, $d=0.3$~D, the dipolar energy scale becomes comparable to $k_BT$ and the dynamics transitions to a semiclassical regime where the elastic cross section is described by the Eikonal approximation \cite{bohn:09},
\begin{align}
\sigma_\mathrm{Eikonal} = \frac{8\pi}{3} d^2\sqrt{\frac{\pi\mu}{2k_BT}},
\label{eq:eikonal}
\end{align}
and the loss rate by classical capture theory,
\begin{align}
\sigma_\mathrm{Capture} = 0.37487 \sqrt{\frac{\pi}{\mu}} 2^{5/6} \Gamma(1/3) d^{4/3} (k_BT)^{-1/6}.
\label{eq:capture}
\end{align}
This cross section derives from a critical impact parameter at which the height of the centrifugal barrier for an isotropic $-C_3 R^{-3}$ interaction coincides with the collision energy,
\begin{align}
b_\ast = \sqrt{3} \left(\frac{C_n}{2E}\right)^{1/3}.
\end{align}
For all impact parameters below $b_\ast$ short-range can be reached classically and these contribute to a cross section $\pi b_{\ast}^2$ \footnote{A factor $1/2$ for the contribution of only odd or even partial waves, rather than all classical impact parameters, cancels against a factor 2 for the loss of two identical molecules}.
In reality, the interaction strength $C_n = -2d^2P_2(\cos\theta)$ is anisotropic,
and we use a sudden approximation to simply average the angular dependence of the cross section, $[-2P_2(\cos\theta)]^{2/3}$, over orientations where this interaction is attractive, which results in the numerical prefactor.
As can be seen, this sudden approximation is not perfectly accurate, but capture theory describes the temperature and dipole dependence well.
For dipole moments below the transition to the classical regime, the dipolar contribution to the cross section is described accurately by the Born approximation \cite{bohn:09},
\begin{align}
\sigma^\mathrm{Bosons}_\mathrm{Born}   &= 1.117 (\mu^2 d^4), \nonumber\\
\sigma^\mathrm{Fermions}_\mathrm{Born} &= 3.351 (\mu^2 d^4),
\label{eq:born}
\end{align}
but the range over which this approximation valid ($d<0.3$~D) while the dipolar length scale is dominant is limited to fermions at low temperatures or bosons where the $s$-wave scattering length happens to be small, which is not shown here as Fig.~\ref{fig:multichannel_T2} shows cross sections averaged over $\delta^s$. 

\begin{figure*}
\begin{center}
\includegraphics[width=\figwidth]{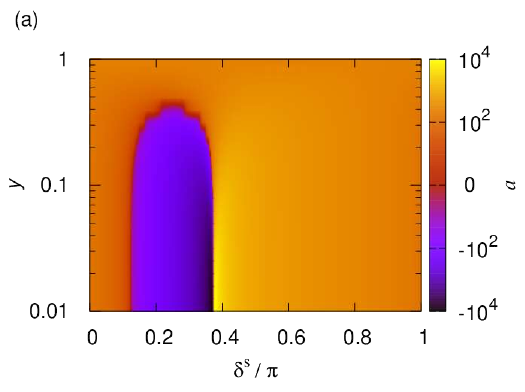}
\includegraphics[width=\figwidth]{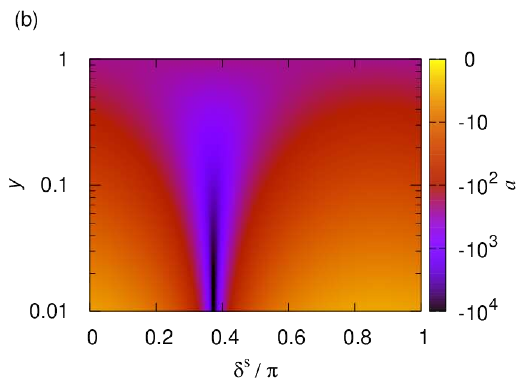}
\includegraphics[width=\figwidth]{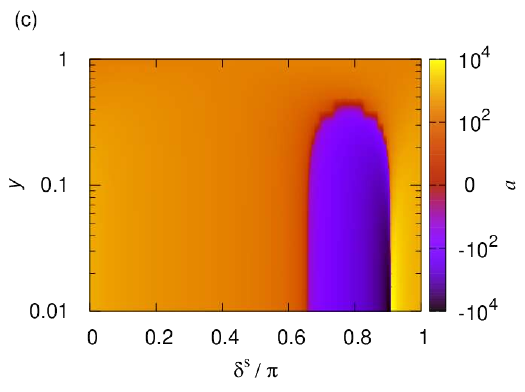}
\includegraphics[width=\figwidth]{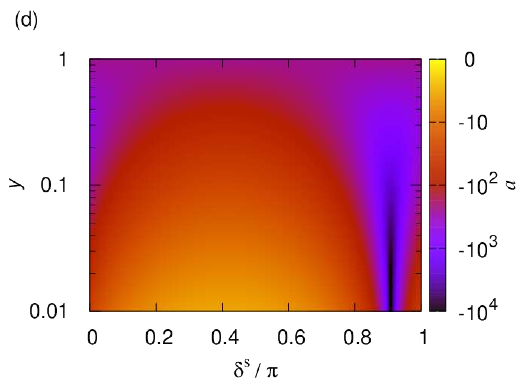}
\caption{ \label{fig:C6} Scattering length as a function of the short-range boundary condition, $\delta^s$ and $y$, for $\efield=0$.
The left and right hand columns show the real and imaginary part of the scattering length, respectively.
The top panels show the analytic result for the pure long-range $R^{-6}$ potential,
and the bottom panels show the results of numerical calculations on the lowest adiabatic potential.
Deviations of the lowest adiabat from its long-range form lead to an additional short-range phase shift,
but otherwise do not affect the scattering length.
}
\end{center}
\end{figure*}

\subsection{Scattering lengths}

In the remainder of this paper we consider the $s$-wave scattering length,
which determines the low-energy scattering behavior for bosons,
as realizable by applying various external fields.
Figure~\ref{fig:C6} shows the $s$-wave ($\ell=0$) scattering length for the rotational van der Waals potential, \emph{i.e.},\ in the absence of an applied field, as a function of the short-range boundary condition, $\delta^s$ and $y$.
The left and right hand columns show the real and imaginary part of the scattering length, respectively.
The top panels show the analytic result for the pure long-range $R^{-6}$ potential, Eq.~\eqref{eq:scatlen}.
The bottom panels show the results of numerical calculations on the lowest adiabatic potential.
This calculation was continued to $R_0=20~a_0$ where the wave functions were matched to the short-range solutions for the $R^{-3}$ short-range potential.
The deviation of the potential from its asymptotic $R^{-6}$ form in this region results in an additional short-range phase shift acquired before the potential reaches its asymptotic form, but apart from this, the scattering lengths are in excellent agreement.
Perhaps unsurprisingly, this confirms numerically that the dynamics is dominated by the $R^{-6}$ long range, and any deviations from this long range form can be described by the short-range phase shift, $\delta^s$.
This \emph{a posteriori} justifies not explicitly including higher-order multipole moments, leave alone modifications of the potential at much shorter range.

\begin{figure*}
\begin{center}
\includegraphics[width=\figwidth]{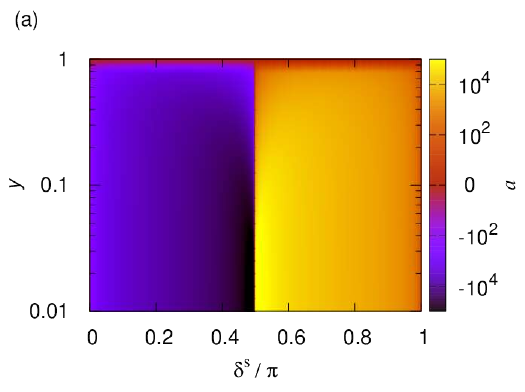}
\includegraphics[width=\figwidth]{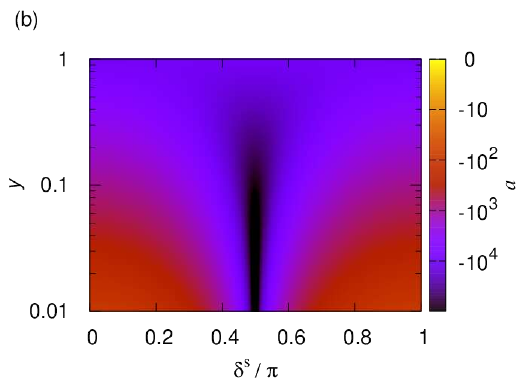}
\includegraphics[width=\figwidth]{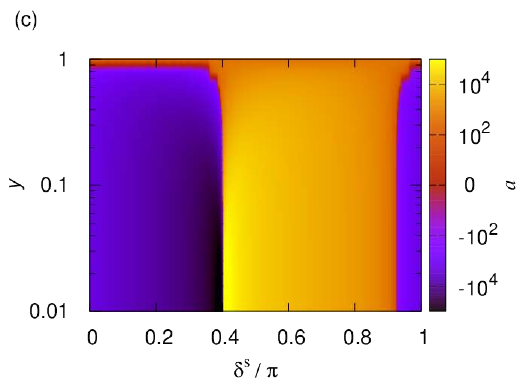}
\includegraphics[width=\figwidth]{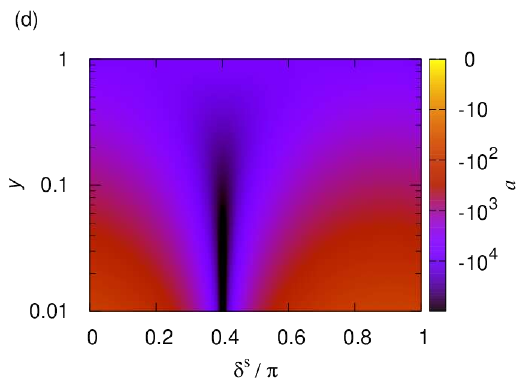}
\caption{ \label{fig:C4}
Scattering length as a function of the short-range boundary condition, $\delta^s$ and $y$, for $\efield=2.7$~kV/cm, inducing a 1~D dipole moment.
The left and right hand columns show the real and imaginary part of the scattering length, respectively.
The top panels show the analytic result for the pure long-range $R^{-4}$ potential,
and the bottom panels show the results of numerical calculations on the lowest adiabatic potential.
Deviations of the lowest adiabat from its long-range form lead to an additional short-range phase shift,
but otherwise do not affect the scattering length.
}
\end{center}
\end{figure*}

Next, we similarly consider the scattering length for a fixed applied electric field.
Figure~\ref{fig:C4} shows the scattering length for $\efield=2.7$~kV/cm, which induces a 1~D dipole moment in the NaK molecules.
The left and right hand columns show the real and imaginary part of the scattering length, respectively.
The bottom panels show numerical results for calculations on the lowest adiabatic potential.
The top panels show the analytic result for the pure long-range $R^{-4}$ potential, Eq.~\eqref{eq:scatlen}.
This again matches closely with the numerical results apart from a difference in the short-range phase shift.

\begin{figure*}
\begin{center}
\includegraphics[width=\figwidth]{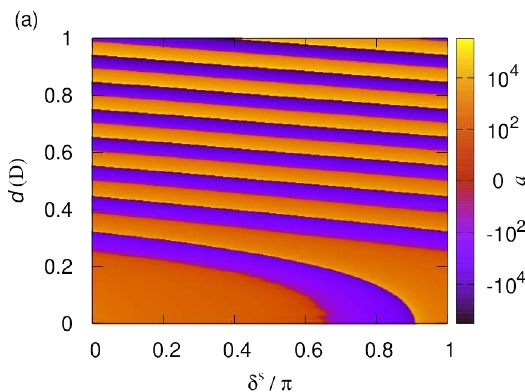}
\includegraphics[width=\figwidth]{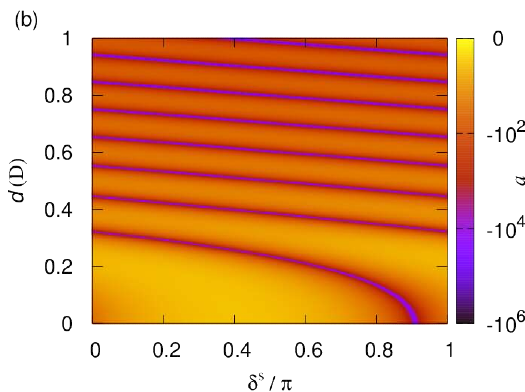}
\includegraphics[width=\figwidth]{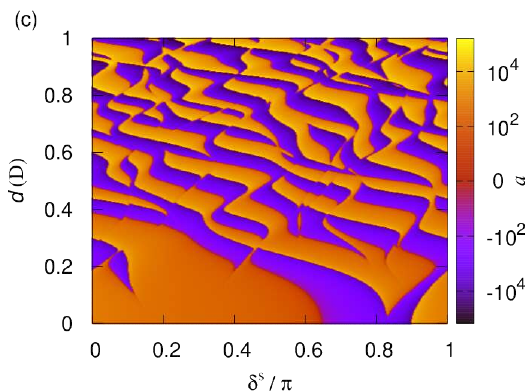}
\includegraphics[width=\figwidth]{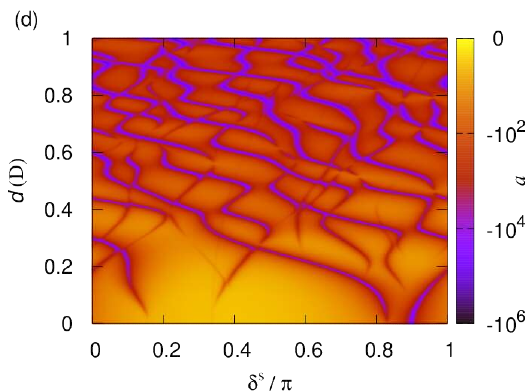}
\caption{ \label{fig:C3}
Scattering length as a function of the short-range phase shift, $\delta^s$, and dipole moment induced by a static electric field for fixed $y=0.01$.
The left and right hand columns show the real and imaginary part of the scattering length, respectively.
The top and bottom panels show the scattering length from single channel and multi-channel calculations, respectively.
}
\end{center}
\end{figure*}

\begin{figure*}
\begin{center}
\includegraphics[width=\figwidth]{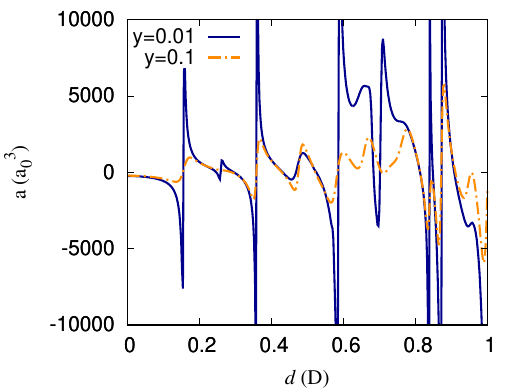}
\caption{ \label{fig:C3lin}
Scattering length as a function of the dipole moment induced by a static electric field for fixed $\delta^s=3\pi/4$.
Resonances that appear as poles in the absence of short-range loss becomes smoother as $y$ is increased.
}
\end{center}
\end{figure*}

\begin{figure*}
\begin{center}
\includegraphics[width=\figwidth]{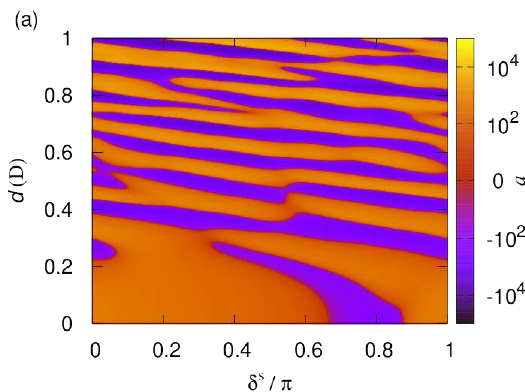}
\includegraphics[width=\figwidth]{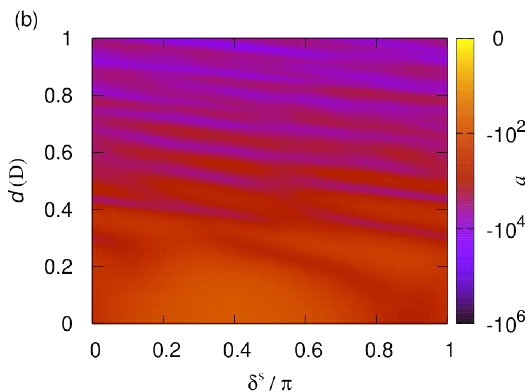}
\caption{ \label{fig:C3quarter}
Scattering length as a function of the short-range phase shift, $\delta^s$, and dipole moment induced by a static electric field for fixed $y=1/4$.
The left and right hand columns show the real and imaginary part of the scattering length, respectively.
}
\end{center}
\end{figure*}

To access the series of dipolar resonances, in what follows we tune the dipolar interactions.
Figure~\ref{fig:C3} shows the scattering length as a function of the dipole moment, $d(\efield)$, induced by applying a static electric field, $\efield$, and the short-range phase shift, $\delta^s$, for fixed $y=0.01$.
Results are shown both for a single channel calculation using the lowest adiabatic potential and for a multi-channel calculation.
In the single channel case, at any given induced moment, the scattering length resembles that of a isotropic $R^{-4}$ potential, as shown above.
This figure shows how the position of the resonances depends on the phase shift, and how new resonances appear with increasing induced dipole moment.
The overall magnitude of the scattering length can be seen to increase as the dipole moment is increased.
In the multi-channel calculation, higher adiabatic potentials contribute further sharper resonances that depend differently on the short-range phase shift,
leading to a complex pattern of crossings.
Figure~\ref{fig:C3lin} illustrates for fixed $\delta^s=3\pi/4$ how the resonances that appear as poles in the absence of short-range loss becomes smoother as $y$ is increased.

Figure~\ref{fig:C3quarter} shows the again scattering length for molecules polarized by a static electric field as a function of the dipole moment, $d(\efield)$, and the short-range phase shift, $\delta^s$ but now for a larger short-range loss parameter $y=1/4$.
It has recently been suggested ultracold molecules may exhibit substantial but nonuniversal loss described by $y=1/4$ \cite{christianen:21}.
Indeed, non-universal short-range loss consistent with this has been observed for RbCs molecules \cite{gregory:19}.
Figure~\ref{fig:C3quarter} shows that in such cases much of the resonance structure is still observable, although the contrast is reduced compared to the case $y=0.01$ shown in Fig.~\ref{fig:C3}.
For a more systematic discussion of the $y$ dependence of the collisional loss rate and elastic cross section we refer the reader back to the previous two subsections.

\begin{figure*}
\begin{center}
\includegraphics[width=\figwidth]{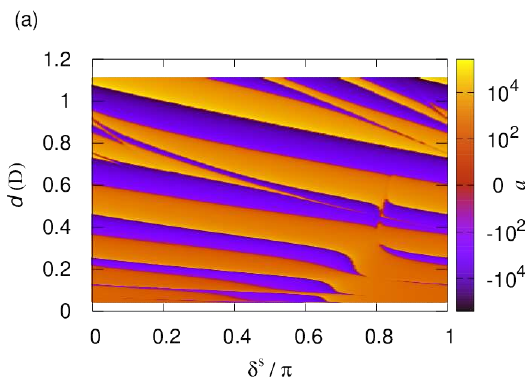}
\includegraphics[width=\figwidth]{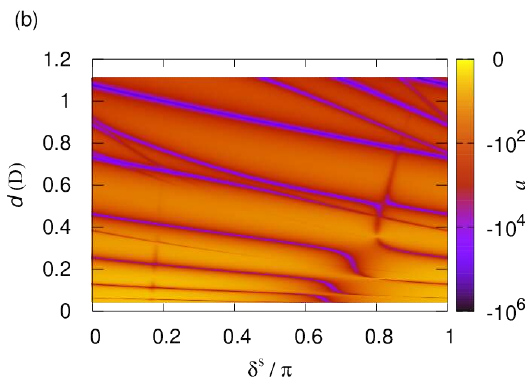}
\includegraphics[width=\figwidth]{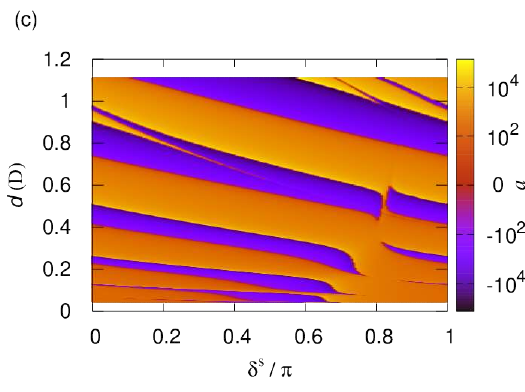}
\includegraphics[width=\figwidth]{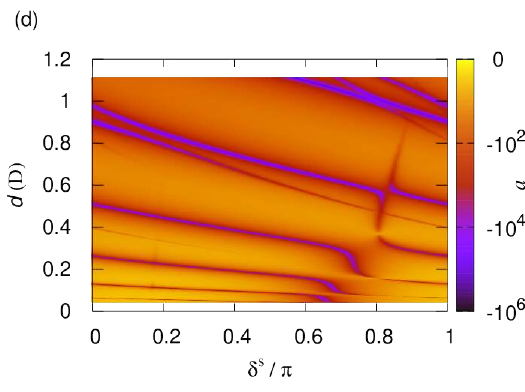}
\caption{ \label{fig:job81}
Scattering length as a function of the short-range phase shift, $\delta^s$, and dipole moment induced by red-detuned microwaves with Rabi frequency $\Omega=2\pi\times 1$~MHz.
The left and right hand columns show the real and imaginary part of the scattering length, respectively.
The top and bottom panels show results for linear and circular polarization, respectively.
}
\end{center}
\end{figure*}

\begin{figure*}
\begin{center}
\includegraphics[width=\figwidth]{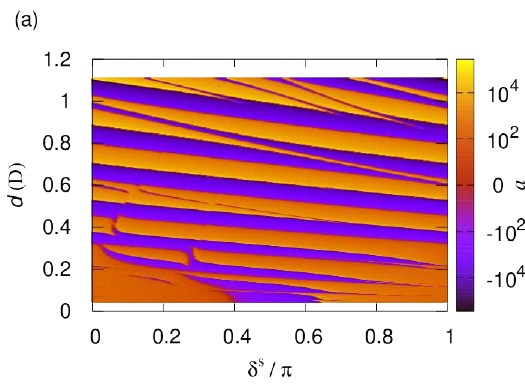}
\includegraphics[width=\figwidth]{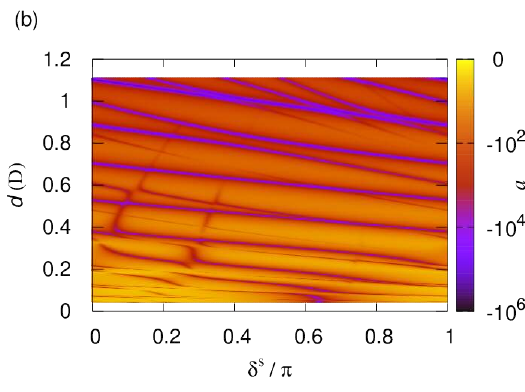}
\caption{ \label{fig:job81x}
Scattering length as a function of the short-range phase shift, $\delta^s$, and dipole moment induced by red-detuned microwaves with Rabi frequency $\Omega=2\pi\times 10$~MHz and linear polarization.
The Rabi frequency is increased relative to that shown in Fig.~\ref{fig:job81}.
As a result, the resonant dipolar interactions at the Condon point are stronger at fixed $\Omega/\Delta$, resulting in the same induced dipole moment but a higher density of scattering resonances.
}
\end{center}
\end{figure*}

Next, we also consider dipolar interactions induced by microwave dressing, rather than applying a static field.
Figure~\ref{fig:job81} shows the scattering length as a function of the short-range phase shift, $\delta^s$, and dipole moment induced by red-detuned microwaves.
This is obtained for fixed $y=0.01$ and Rabi frequency $\Omega=2\pi\times 1$~MHz.
The resulting scattering length shows some similarities to that obtained for static electric fields, Fig.~\ref{fig:C3}.
In particular, we observe a set of tunable resonances on top of background cross sections which increase with the induced dipole moment. 
However, the density of resonance is lower than that obtained for static electric fields, and hence is not explained by the first-order dipolar interactions.
This occurs because the interactions are dominated by resonant dipolar interactions \cite{karman:21a}.
For off-resonant dressing, for small induced dipole moments, this can be understood as a crossing between the bare initial state and a resonantly interacting excited state that is avoided by Rabi coupling between the two.
The range over which resonant dipolar interactions are tuned for a fixed range of induced dipole moment, $d$, increases with $\Omega$,
and hence the density of resonances increases with $\Omega$ as is shown in Fig.~\ref{fig:job81x}.

\begin{figure*}
\begin{center}
\includegraphics[width=\figwidth]{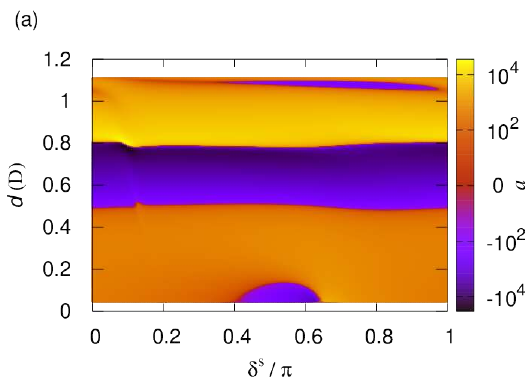}
\includegraphics[width=\figwidth]{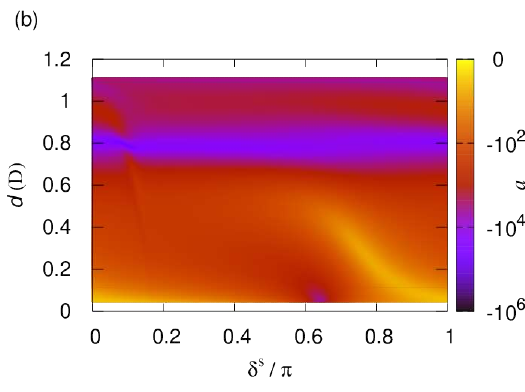}
\includegraphics[width=\figwidth]{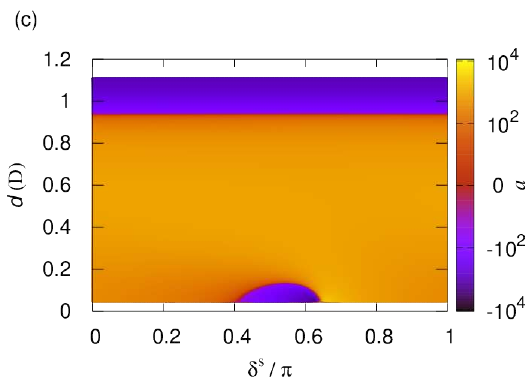}
\includegraphics[width=\figwidth]{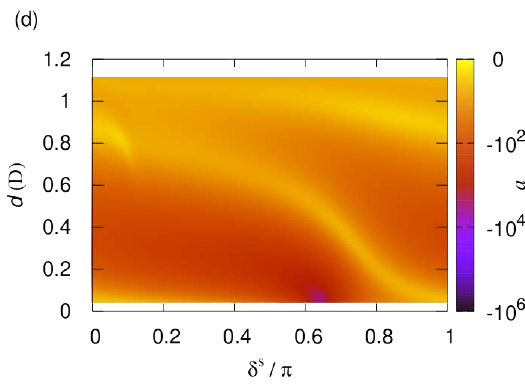}
\caption{ \label{fig:job82}
Scattering length as a function of the short-range phase shift, $\delta^s$, and dipole moment induced by blue-detuned microwaves with Rabi frequency $\Omega=2\pi\times 1$~MHz.
The left and right hand columns show the real and imaginary part of the scattering length, respectively.
The top and bottom panels show results for linear and circular polarization, respectively.
}
\end{center}
\end{figure*}

Then we change the sign of the detuning from the rotational transition.
Figure~\ref{fig:job82} shows the scattering length as a function of the short-range phase shift, $\delta^s$, and dipole moment induced by blue-detuned microwaves.
This is obtained for fixed $y=0.01$ and Rabi frequency $\Omega=2\pi\times 1$~MHz.
Contrary to the case for red detuning, the resulting scattering length does not resemble that obtained for static electric fields.
This occurs as, for blue detuning, the short-range modifications of the interaction potential are repulsive rather than attractive \cite{karman:21a}.
As a result, no significant flux reaches short range, and the scattering becomes independent of the short-range phase shift.
For resonant dressing with circularly polarization, this realizes microwave shielding \cite{karman:18d,karman:19c,karman:20}, and the imaginary part of the scattering length becomes small.
For linear polarization, shielding is ineffective due to nonadiabatic transitions outside the short-range repulsive regions, and the imaginary part of the scattering length remains large.

\begin{figure*}[h!]
\begin{center}
\includegraphics[width=\figwidth]{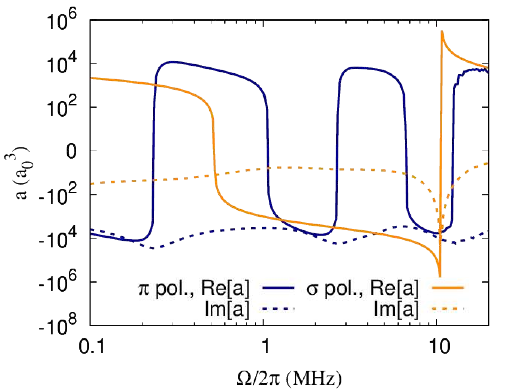}
\caption{ \label{fig:job72}
Scattering length as a function of the Rabi frequency, $\Omega$, for resonant dressing ($\Delta=0$) with blue-detuned microwaves.
Results are shown for $y=0.01$ and $\delta^s=0$, whereas the results are essentially independent of $\delta^s$ as shown in Fig.~\ref{fig:job82}.
}
\end{center}
\end{figure*}

Finally, we consider also for blue detuning the dependence on the intensity of the microwaves,
parameterized by the Rabi frequency, $\Omega$.
Figure~\ref{fig:job72} shows the scattering length as a function of $\Omega$ for resonant dressing with blue-detuned microwaves.
Results are shown for $y=0.01$ and $\delta^s=0$, whereas the results are essentially independent of $\delta^s$ as shown in Fig.~\ref{fig:job82}.
By increasing the Rabi frequency one moves the repulsive shield to shorter $R$, as this emerges where the resonant dipole-dipole interaction is dominant over $\hbar\Omega$.
The interaction potential on the outside of the repulsive shield is then deepened.
As a result, a series of resonances emerges which has previously been suggested can be used to control the scattering length while shielding from losses \cite{lassabliere:18}.

\section{Energy dependence of the short-range boundary condition}

In this paper we have presented a method for imposing general short-range boundary conditions -- in the spirit of quantum defect theory -- in coupled-channels calculations of collisions between ultracold molecules.
The coupled-channels calculations treat the long-range interactions between the molecules exactly,
and parameterized the short range by a phase shift $\delta^s$ and loss parameter $y$.
Throughout the paper, to illustrate the method, we have imposed short-range boundary conditions that are independent of the applied external fields,
the collision energy,
and the centrifugal angular momentum.
This energy and angular momentum insensitive boundary condition has been applied successfully to atomic collisions \cite{gao:01}.
However, it is not completely clear that this boundary condition applies also to ultracold molecular collisions.
In particular, Mayle \emph{et al.} have shown that the density of states of molecule-molecule collision complexes may lead to a highly energy dependent short-range phase shift \cite{mayle:12,mayle:13}
\begin{align}
\tan\delta^s = - \sum_\nu \frac{\gamma_\nu}{E-E_\nu},
\end{align}
where $E_\nu$ are the positions of a dense set of resonances,
with density $\rho$,
and the elastic widths are of order $1/2\pi\hbar\rho$.
Christianen \emph{et al.} have suggested an extension that incorporates short-range loss \cite{christianen:21}.

The approach developed here can also be used with an energy dependent boundary condition once $\delta^s(E)$ and $y(E)$ are known, for example from the models of Refs.~\cite{mayle:12,christianen:21}.
In fact, the present method is highly suitable for calculations with highly energy dependent boundary conditions since the boundary conditions are imposed after both sets of linearly independent solutions to the scattering problem are determined.
The step of imposing different boundary conditions can be repeated at almost no computational cost without repeating the computer intensive coupled-channels scattering calculations,
leading to a large speed-up compared to calculations where one initializes the short-range boundary condition and subsequently propagates the solution \cite{wang:15}, \emph{i.e.}, a process that needs to be repeated for each boundary condition that one would like to impose.

\section{Conclusions \label{sec:conclusions}}

In this work, we have presented an efficient method for performing multichannel quantum defect scattering calculations.
The main advantage is that the scattering calculation itself need not be repeated to impose various boundary conditions,
which are parameterized here by a short-range loss parameter, $y$, and short-range phase shift, $\delta^s$.
We give explicit expressions for the short-range reference functions, required to impose these boundary conditions,
for arbitrary $R^{-n}$ short-range potentials in two approximations;
neglect of the local adiabat's channel energy or a WKB approximation that does account for the local adiabat's channel energy.
The $R^{-n}$ form of the potential here refers only to the small $R$ behavior, whereas essentially arbitrary multichannel interactions at long range are treated numerically.
This is illustrated here by application to collisions of ultracold NaK molecules in external static and microwave fields,
which leads to multichannel collisions with adiabatic potentials between $R^{-3}$ and $R^{-6}$.

Our interest here has been dipolar collisions between ultracold molecules.
The dipole-dipole interaction can be tuned to cause resonances if the potential supports bound states, which is not the case for universal short-range loss, $y=1$.
Here, we studied how a series of dipolar resonances becomes observable as short-range losses are eliminated, for example in repulsive box potentials \cite{christianen:19a,yan:20}
We find the series emerges for short-range losses as high as $y=0.5$, which means the series could already be observable for RbCs \cite{gregory:19}.
It has recently been suggested ultracold molecules may ubiquitously exhibit substantial but nonuniversal loss described by $y=1/4$ \cite{christianen:21}, as observed for RbCs,
which would render the resonances studied here observable for many molecules.
Another avenue along which tunable long-range dipolar interactions can be realized in the absence of short-range losses is microwave shielding \cite{karman:18d,lassabliere:18,anderegg:21,schindewolf:22}.
A large part of the series of resonances is observable at experimentally realized temperatures around $1~\mu$K.

\section{Acknowledgement}

The author gratefully acknowledges stimulating discussions with Martin Zwierlein.

\end{document}